\documentclass[aps,amsmath,amssymbols,superscriptaddress,twocolumn]{revtex4}
\usepackage[]{graphicx}
\usepackage{color}
\usepackage{bbold}
\usepackage{braket}
\usepackage[colorlinks=true, pdfstartview=FitV, linkcolor=blue, citecolor=red, urlcolor=black]{hyperref}
\usepackage{orcidlink}
\usepackage{pst-node}
\usepackage{tikz}
\newcommand{\be}{\begin{equation}}
\newcommand{\ee}{\end{equation}}
\newcommand{\ben}{\begin{eqnarray}}
\newcommand{\een}{\end{eqnarray}}
\newcommand{\bes}{\begin{subequations}}
\newcommand{\ees}{\end{subequations}}
\def\bal#1\eal{\begin{align}#1\end{align}}
\usepackage{comment}

\newcommand{\bfi}{\begin{figure}}
\newcommand{\efi}{\end{figure}}
\newcommand{\bc}{\begin{center}}
\newcommand{\ec}{\end{center}}
\newcommand{\sech}{\mbox{sech}}

\begin{document}
    \title{Geometrically constrained multifield models with BNRT solutions}
        
        \author{M.A. Marques\,\orcidlink{0000-0001-7022-5502}}
        \email[]{marques@cbiotec.ufpb.br}
    \affiliation{Departamento de Biotecnologia, Universidade Federal da Para\'iba, 58051-900 Jo\~ao Pessoa, PB, Brazil}
    
    \author{R. Menezes\,\orcidlink{0000-0002-9586-4308}}
     \email[]{rmenezes@dcx.ufpb.br}
    \affiliation{Departamento de Ci\^encias Exatas, Universidade Federal da Para\'iba, 58297-000 Rio Tinto, PB, Brazil}

\begin{abstract}
In this paper, we investigate multifield models in which the two-field BNRT model is coupled to a third field through mediator functions in the Lagrangian density. To conduct the investigation, we obtain the equations of motion and develop a first-order formalism based on energy minimization. Two possibilities are considered: i) the third field acting in the mediator functions to modify the BNRT solutions; ii) the BNRT fields feeding the mediator function to produce effects in the kink solution of the third field. In the case i), the results show that the solutions may be related to the standard ones with the coordinate redefined in terms of the mediator functions if they are equal. This allows to induce effects similar to geometric constrictions in the core, or to compactify the tail of the BNRT solutions. If the mediator functions differ one from another, we show that the effects are distinct, with the compactification of just one of the two-field solutions. In the case ii), the orbit parameter of the model plays an important role, modifying the mediator function that induces changes in the profile of the kink associated to the third field.
\end{abstract}
\maketitle

\section{Introduction}
Scalar field models are of interest in Field Theory as they may be applied in several branches of Physics. In particular, they can be used in the study of topological structures. The simplest model which support these objects is described by a Lagrangian density of a single real scalar field in $(1,1)$ spacetime dimensions. In this situation, one finds the so-called kinks, which are static solutions of the equations of motion connecting two neighbor minima of the potential \cite{vachaspati}. These structures may appear, for instance, in the investigation of magnetic materials \cite{mag1,mag2}, where they can describe specific behavior, or in the curved spacetime scenario, where they give rise to thick branes \cite{RS,dewolfe,gremm1}.

Models of a single scalar field are useful in the modeling of N\'eel walls separating two magnetic domains. Another type of domain wall that is of interest is the Bloch wall, which engenders two degrees of freedom. From the point of view of Field Theory, one may accommodate the extra degree freedom by including an additional scalar field in the Lagrangian density, so one must work with two scalar fields. Since the equations of motion are nonlinear, the investigation becomes harder. In this direction, a two-field model with analytical results was introduced in Refs.~\cite{bnrt1,bnrt2} by Bazeia {\it et al}, which is known as the BNRT model \cite{izquierdoorbit}. The solutions are compatible with a first-order formalism in the lines of the Bogomol'nyi procedure \cite{bogo}, ensuring that the minimum energy of the system is attained.

The localized solutions found in Refs.~\cite{bnrt1,bnrt2} were used in several contexts over the years \cite{shifman,bba,adiabatic1,adiabatic2,alvaro,izquierdoquantum,bnrtlorentz1,bnrtlorentz2,bnrtlorentz3,bnrtlorentz4,bnrtadam,bnrtavelino,blochbrane,bnrtbrane1,bnrtbrane2,bnrtbrane3,bnrtbrane4,bnrtbrane5,gani1}. In particular, in scenarios with violation of Lorentz symmetry, where the solutions may become non monotonic \cite{,bnrtlorentz1,bnrtlorentz2,bnrtlorentz3,bnrtlorentz4}, in collisions of domain walls \cite{gani1}, where an effect similar to elastic reflection with some delay may emerge, in the investigation of spectral walls, which may appear as singularities of the moduli space \cite{bnrtadam}, and in the study of bifurcation and pattern changing in domain wall networks \cite{bnrtavelino}. The BNRT model was also considered in the braneworld scenario with a single extra dimension of infinite extent, giving rise to the Bloch brane \cite{blochbrane}. In this case, the thick brane may support an internal structure depending on the values of the parameters associated to the orbit. It has also been considered in the study of localization of fermion \cite{bnrtbrane2,bnrtbrane3} and gauge fields \cite{bnrtbrane3,bnrtbrane4,bnrtbrane5}. Other class of two-field models was proposed in Ref.~\cite{twofield2}, where the interaction occurs with a periodic potential for a field and a polynomial for the other one, leading to composite-kink solutions.

Multifield models with three scalar fields were previously considered in Refs.~\cite{three1,three2,three3}. The presence of a third field makes the task of obtaining analytical results even harder, as one must solve three differential equations to find the set of orbits. However, recently, an interesting possibility, in which the first-order equation that governs one field can be decoupled from the others was presented in Refs.~\cite{const1,constbloch,const2,const3}. In the two-field models investigated in Refs.~\cite{const1,const2}, one field is used to modify the internal structure of kinks, giving rise to a multikink profile with several plateaus. Interestingly, this mechanism seems to simulate the presence of geometric constrictions in domain walls \cite{mag1}, modifying the slope of the kink at its center. Later, we have developed this procedure to modify the tail of kinks, which may become long range or compact \cite{const3}. This mechanism was extended to a three-field model that allows for the induction of internal structure in the solutions of the elliptic orbits in the BNRT model in Ref.~\cite{constbloch}. The possibility of investigating multifield models engendering triplets was considered in Ref.~\cite{gani2,gani3}; in this case, it was shown that the presence of degrees of freedom localized on the wall depends on the parameters of the model.

In this work, we study multifield models involving the BNRT model with the presence of an additional scalar field. In Sec.~\ref{sec1}, we consider a scalar field to modify the kinetic terms of the two fields associated to the BNRT model via two mediator functions in the Lagrangian density. In this scenario, we explore the case in which the mediator functions are equal, allowing for the introduction of a novel geometric coordinate to become the argument of the standard minimum-energy solutions; or different, when we cannot relate the solutions to the standard ones and the problem requires a numerical approach. The results show that two-field solutions with compact support may emerge and, for specific conditions in the mediator functions, only one of the two degrees of freedom associated to the domain wall presents significant effects. As one knows, the study of compactons started in Ref.~\cite{rosenau} and, as far as we know, this is the first time that compact solutions were obtained in multifield models. We have also found that, if the mediator functions are not equal, intersection points may appear in the orbits. In Sec.~\ref{sec2}, we investigate a second possibility, in which we invert the logic and consider that the BNRT fields act to modify the third scalar field through a mediator function. In both cases, natural units and dimensionless fields and coordinates are taken. We obtain the equations of motion and develop a first-order formalism based on the minimization of energy. We also find the equation of the orbits associated to the BNRT fields and obtain the conditions under which the orbits can be found. The investigation ends in Sec.~\ref{secend}, where we present some final comments and perspectives for future research. 
\section{Model 1}\label{sec1}
We start our investigation considering a three-field model in $(1,1)$ flat spacetime dimensions:
\be\label{lmodel1}
\begin{aligned}
    \mathcal{L} &= \frac12f(\psi)\partial_\mu\phi\partial^\mu\phi + \frac12g(\psi)\partial_\mu\chi\partial^\mu\chi \\
    &+\frac12\partial_\mu\psi\partial^\mu\psi - V(\phi,\chi,\psi),
\end{aligned}
\ee
where $\phi$, $\chi$ and $\psi$ are real scalar fields, $f(\psi)$ and $g(\psi)$ are non-negative functions that modify the dynamical terms of $\phi$ and $\chi$, and $V(\phi,\chi,\psi)$ is the potential. In Ref.~\cite{constbloch}, this model was investigated in the special case $f(\psi)=g(\psi)$, using a first-order framework based on energy minimization, which allows for the field $\psi$ being governed by a first-order equation decoupled from the other ones. In this specific situation, it was shown that $\psi$ can be used to control the internal structure of the Bloch wall solution described by $\phi$ and $\chi$. In the current work, we explore the general case, seeking novel configurations in the BNRT model \cite{bnrt1,bnrt2,izquierdoorbit}. Below, we display a scheme to show how the field $\psi$ is coupled to the other fields, $\phi$ and $\chi$, through the mediator functions $f(\psi)$ and $g(\psi)$.
\begin{center}
\tikzset{terminal2/.style  = {draw, circle, color=white, text=black, fill=white, minimum size=2em}}
\begin{tikzpicture}[
terminal/.style={ circle,     minimum width=1cm,       minimum height=1cm,      ultra thin, draw=black,
       font=\itshape}
      ]
\matrix[row sep=-0.2cm,column sep=0.2cm] {%
       &\node [terminal2](p4) {$f(\psi)$};& \node [terminal](p2) {$\phi$};& \\
    \node [terminal](p1) {$\psi$};    &&  & \\
     & \node [terminal2](p4) {$g(\psi)$}; &  \node [terminal](p3) {$\chi$};&\\
}; 

\draw[gray]   (p1) edge [->,>=stealth,shorten <=2pt, shorten >=2pt,thick] (p2)
        (p1)  edge [->,>=stealth,shorten <=2pt, shorten >=2pt, thick] (p3);
\end{tikzpicture}
\end{center}
By varying the action associated to \eqref{lmodel1} with respect to $\phi$ and $\chi$, one gets
\be
\partial_\mu\left(f(\psi)\,\partial^\mu \phi\right)+V_\phi=0 \quad\text{and}\quad \partial_\mu\left(g(\psi)\,\partial^\mu \chi\right) + V_\chi=0,
\ee
where $V_\phi = \partial V/\partial\phi$ and $V_\chi = \partial V/\partial\chi$. The field $\psi$ is governed by
\be
\partial_\mu\partial^\mu\psi -\frac{1}{2} f_\psi \partial_\mu\phi\partial^\mu\phi - \frac{1}{2}g_\psi \partial_\mu\chi\partial^\mu\chi + V_\psi =0,
\ee
where $f_\psi = df/d\psi$, $g_\psi = dg/d\psi$ and $V_\psi = \partial V/\partial\psi$. Since we are interested in topological solutions, we take static configurations. In this case, the equations of motion become 
\be
\left(f(\psi)\, \phi'\right)'=V_\phi \quad\text{and}\quad
\left(g(\psi)\, \chi'\right)' = V_\chi
\ee
and
\be
\psi'' -\frac{1}{2} f_\psi  {\phi'}^2 - \frac{1}{2}g_\psi {\chi'}^2 = V_\psi.
\ee
The prime represents the derivative with respect to $x$. These equations must be solved to get localized structures, so the fields must connect two minima of the potential, which depend on the model under investigation, asymptotically. The energy density can be calculated standardly; it is
\be\label{rhogeral}
\rho(x)=\frac12f(\psi){\phi'}^2+\frac12g(\psi){\chi'}^2 + \frac12{\psi'}^2 + V(\phi,\chi,\psi).
\ee
By integrating the above expression, one gets the energy. In Ref.~\cite{constbloch}, first-order equations were obtained in the case $f(\psi)=g(\psi)$. Since we are considering the general case, in which the aforementioned equality may not be attained, we now develop the Bogomol'nyi procedure \cite{bogo} for it. By introducing an auxiliary function $W(\phi,\chi,\psi)$, the energy density can be rewritten as
\be
\begin{aligned}
    \rho(x)&=\frac12f(\psi)\left(\phi'\mp \frac{W_\phi}{f(\psi)} \right)^2+ \frac12g(\psi)\left(\chi'\mp \frac{W_\chi}{g(\psi)} \right)^2 \\
    &\hspace{4mm}+ \frac12\left(\psi'\mp W_\psi \right)^2 + V(\phi,\chi,\psi) \\
   & \hspace{4mm}- \frac12\left(\frac{W^2_\phi}{f(\psi)} + \frac{W^2_\chi}{g(\psi)} + W_\psi^2\right) \pm W'.
\end{aligned}
\ee
By taking the potential in the form 
\be\label{potw}
V(\phi,\chi,\psi) = \frac12\frac{W^2_\phi}{f(\psi)} + \frac12\frac{W^2_\chi}{g(\psi)} + \frac12W^2_\psi,
\ee
the energy density becomes
\be
\begin{aligned}
    \rho(x)&=\frac12f(\psi)\left(\phi'\mp \frac{W_\phi}{f(\psi)} \right)^2+ \frac12g(\psi)\left(\chi'\mp \frac{W_\chi}{g(\psi)} \right)^2 \\
    &\hspace{4mm}+ \frac12\left(\psi'\mp W_\psi \right)^2 \pm W'.
\end{aligned}
\ee
This shows that the energy has a bound,
\be\label{eb1}
\begin{aligned}
    E\geq E_B &= |W(\phi(+\infty),\chi(+\infty),\psi(+\infty))\\
              &\hspace{5mm}-W(\phi(-\infty),\chi(-\infty),\psi(-\infty))|.
\end{aligned}
\ee
The energy is minimized for solutions obeying the first-order equations $\psi'=\pm W_\psi$, $\phi'= \pm W_\phi/f(\psi)$ and $\chi'= \pm W_\chi/g(\psi)$. Since we are interested in taking $\psi$ as a source of modifications in the other fields, we consider that it is decoupled in the auxiliary function, which we take as 
\be\label{wpq}
W(\phi,\chi,\psi) = P(\phi,\chi) + Q(\psi).
\ee
In this case, the potential in Eq.~\eqref{potw} takes the form
\be\label{vpq}
V(\phi,\chi,\psi) = \frac12\frac{P^2_\phi}{f(\psi)} + \frac12\frac{P^2_\chi}{g(\psi)} + \frac12Q^2_\psi.
\ee
The first-order equation that governs the source field, $\psi$, gets the form
\be\label{fopsi}
\psi'=\pm Q_\psi.
\ee
We want to study solutions with topological nature, so we impose that this field connects two values asymptotically with null derivative, i.e., $\psi(\pm\infty)=\psi_\pm$ and $\psi'(\pm\infty)\to0$. This makes the contribution $Q_\psi^2$ in the potential \eqref{vpq} vanish asymptotically independently from the other fields.
    
The equation \eqref{fopsi} allows us to calculate the solution $\psi(x)$ and use it in the other first-order equations, which become
\be\label{foblochfg}
\phi'= \pm\frac{P_\phi}{f(\psi(x))} \quad\text{and}\quad \chi'= \pm \frac{P_\chi}{g(\psi(x))}.
\ee
These equations are coupled. In order to find topological configurations, we impose that the pair $(\phi,\chi)$ connect minima of the potential \eqref{vpq}. Notice that the solutions of the above equations do not depend on the function $Q_\psi$, so the minima connected by the solution $(\phi,\chi)$ have no relation to the ones connected by $\psi$. We remark that the first-order equations \eqref{fopsi} and \eqref{foblochfg} with upper sign can be related to the ones with lower sign by the change $x\to-x$. So, in the rest of this section, we only consider the upper sign.

The energy density in Eq.~\eqref{rhogeral} can be written as the sum of two contributions, in the form $\rho(x) = \rho_1(x) + \rho_2(x)$, where
\bes\label{rhofo}
\begin{align} \label{rhophichi}
    &\rho_1(x) = \frac{P^2_\phi}{f(\psi(x))} + \frac{P^2_\chi}{g(\psi(x))},\\ \label{rhopsi}
    &\rho_2(x) = Q_\psi^2.
\end{align}
\ees
By integrating these expressions, one can show that each contribution in the energy is $E_1 = |P_+-P_-|$ and $E_2 = |Q_+-Q_-|$, where $P_\pm = P(\phi(\pm\infty),\chi(\pm\infty))$ and $Q_\pm = Q(\psi(\pm\infty))$. This matches with the expression obtained in Eq.~\eqref{eb1}.

To feed the mediator functions $f(\psi)$ and $g(\psi)$ in the multifield model \eqref{lmodel1} in the minimum energy regime, we consider
\be\label{qpsi4}
Q(\psi) = \alpha\psi-\frac{\alpha}{3}\psi^3,
\ee
where $\alpha$ is a positive-real parameter. This function leads to a contribution $\alpha^2(1-\psi^2)^2/2$ in the potential \eqref{vpq} due to the term $Q_\psi^2/2$. From Eq.~\eqref{fopsi}, we get the solution $\psi(x) = \tanh(\alpha x)$, whose associated energy density is $\rho_2 = \alpha^2\sech^4(\alpha x)$. By integrating the last expression, one gets $E_2 = 4\alpha/3$, as expected from the Bogomol'nyi procedure below Eq.~\eqref{rhofo}. 

In the case $f(\psi)=g(\psi)=1$, in which $\psi$ does not affect $\phi$ and $\chi$, Eqs.~\eqref{foblochfg} can be used to get an orbit that relates $\phi$ and $\chi$ in some situations \cite{izquierdoorbit}. In the general case, in which $f(\psi)$ and $g(\psi)$ are present, Eqs.~\eqref{foblochfg} lead to
\be\label{orbiteq}
\frac{d\phi}{d\chi} = \frac{g(\psi(x))}{f(\psi(x))}\frac{P_\phi}{P_\chi}.
\ee
The explicit dependence on the coordinate $x$ in the above expression makes the task of finding the orbit be impossible, as the orbit equation must depend exclusively on the fields. However, when $f(\psi)=g(\psi)$, this issue disappear, such that the orbit equation becomes $d\phi/d\chi = P_\phi/P_\chi$.

An interesting model that couples two fields is the BNRT, which was introduced in Refs.~\cite{bnrt1,bnrt2}. It is given by
\be\label{pbnrt}
P(\phi,\chi) = \phi - \frac13 \phi^3 - r \phi \chi^2,
\ee
where $0<r<1$ is a parameter that controls the strength of the coupling between the fields. The derivatives of this function are $P_\phi = 1-\phi^2-r\chi^2$ and $P_\chi = -2r\phi\chi$. Since the potential \eqref{vpq} is non negative, its minima must obey $P_\phi/f(\psi)=0$ and $P_\chi/g(\psi)=0$. Thus, supposing that $f(\psi)$ and $g(\psi)$ does not lead to divergences stronger than $P_\phi$ and $P_\chi$, the minima of the potential are 
\be
v_{h\pm} = (\pm 1,0) \quad\text{and}\quad v_{v\pm} = (0,\pm1/\sqrt{r}),
\ee
considering each minimum defined by the pair $(\phi,\chi)$. So, the potential engender a set of four minima located in the vertices of a diamond, with $v_{h\pm}$ denoting the horizontal and $v_{v\pm}$ standing for the vertical ones. There are five topological sectors. One sector connects the minima $v_{h\pm}$, supporting an infinite number of solutions with energy $E_1 = 4/3$. Each one of the other four sectors that connect neighbor minima engender a single solution with energy $E_1 = 2/3$.

\begin{figure}[t!]
    \centering
    \includegraphics[width=4.2cm]{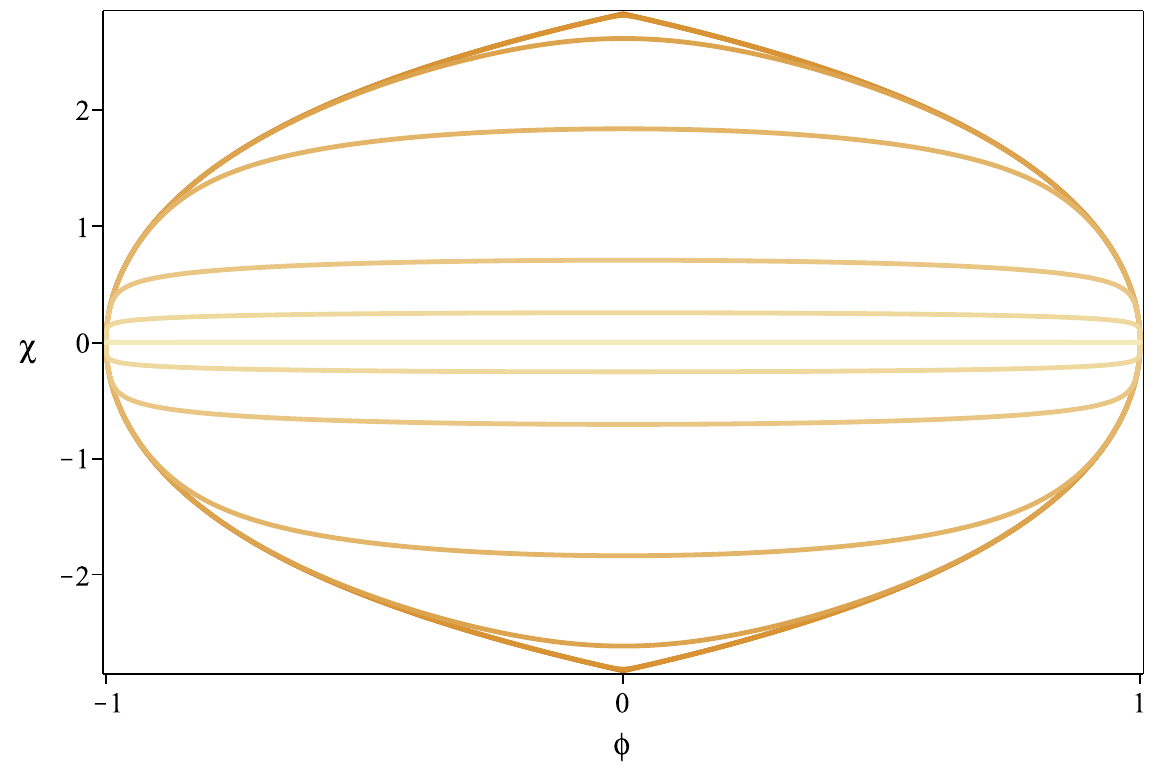}
    \includegraphics[width=4.2cm]{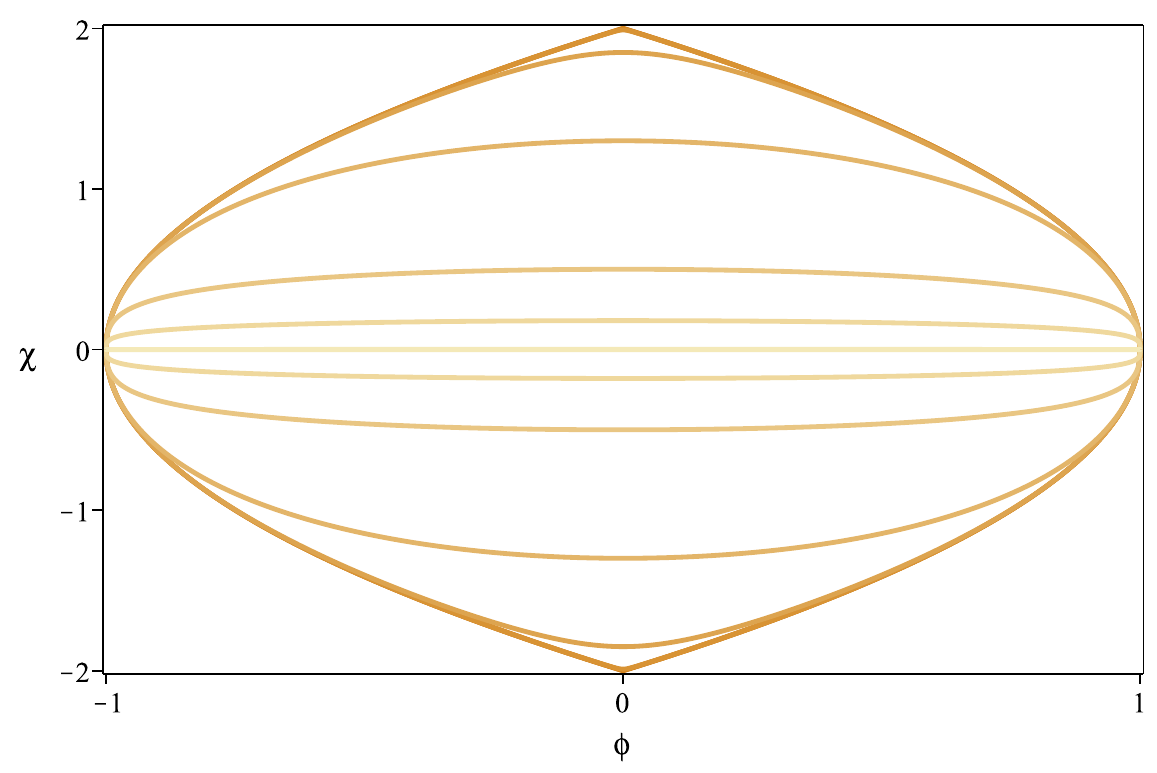}
    \includegraphics[width=4.2cm]{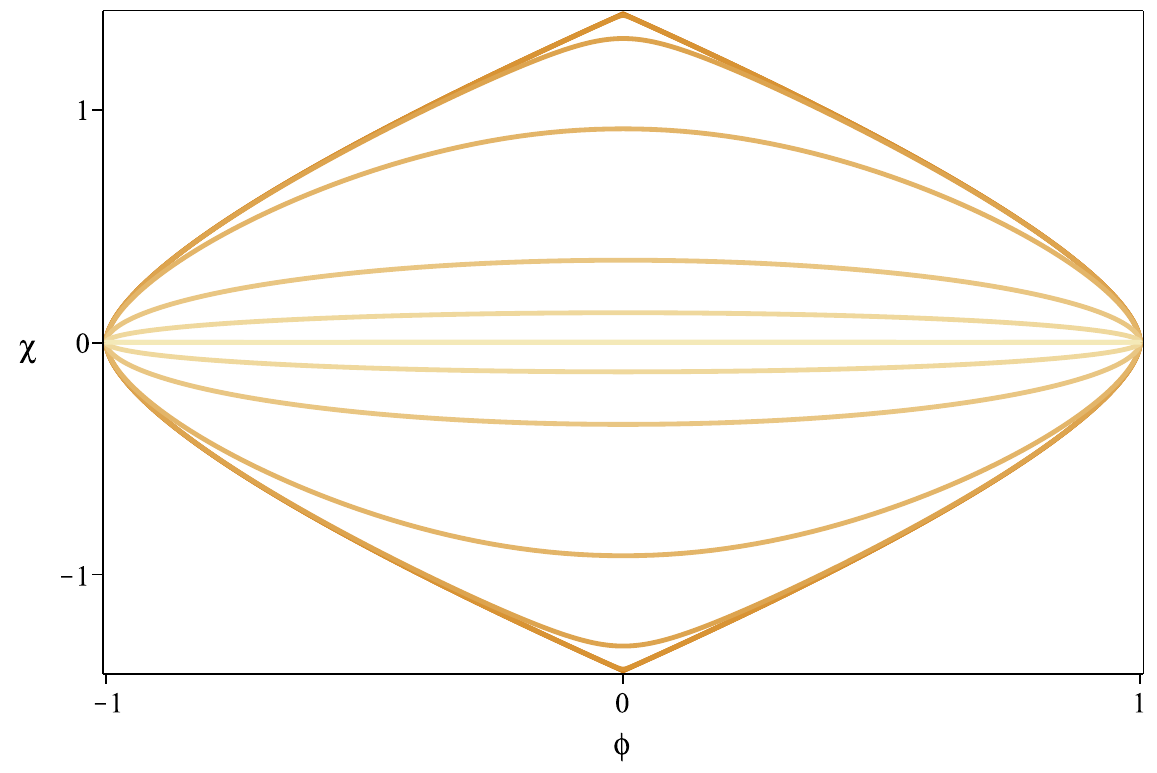}
    \includegraphics[width=4.2cm]{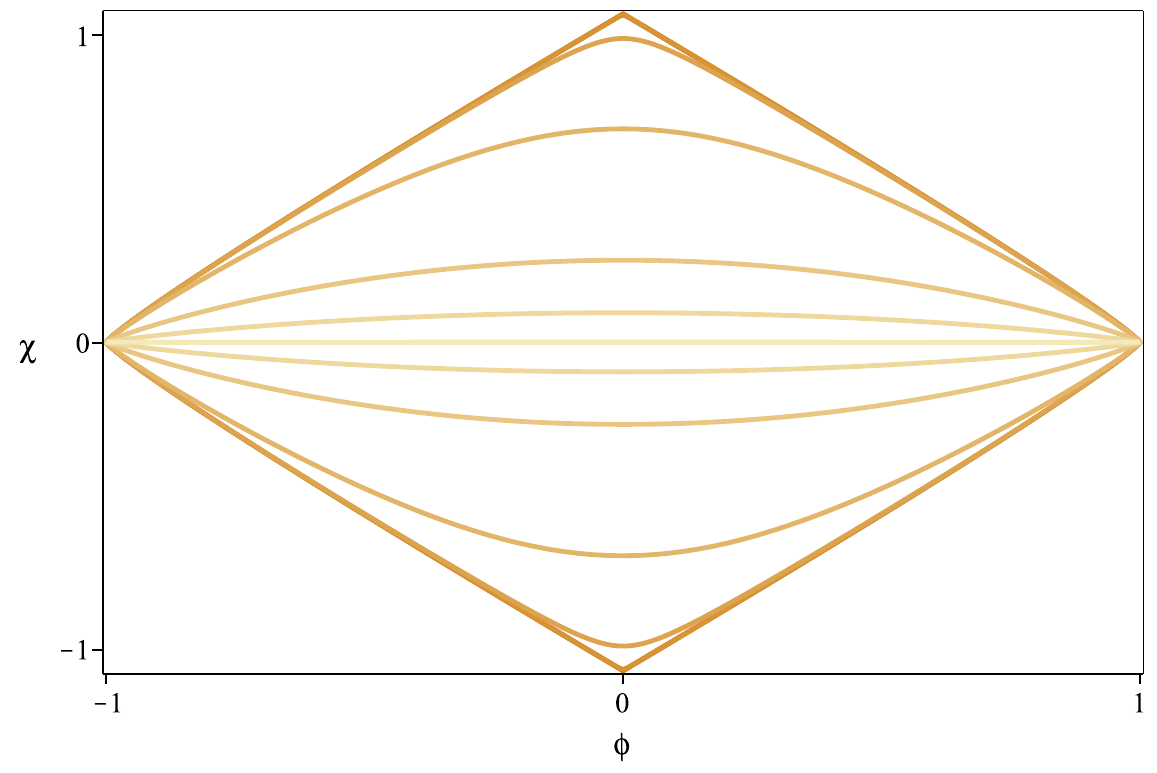}
    \caption{The orbits in Eq.~\eqref{bnrtorbitr} for the BNRT model. We have taken $r=1/8$ (top left), $1/4$ (top right), $1/2$ (bottom left) and $7/8$ (bottom right), with several values of $C$.}
    \label{figorbit}
\end{figure}

For this model, the equation \eqref{orbiteq} with $f(\psi)=g(\psi)$ becomes
\be
\frac{d\phi}{d\chi}=-\frac{1-\phi^2-r\chi^2}{2r\phi\chi}.
\ee
One can use an integrating factor to find the general orbit that connects the minima of the potential. It has the form
\be\label{bnrtorbitr} 
\phi^2 = \begin{cases} \displaystyle
    1-\frac{r}{1-2r}\chi^2 + C \chi^{\frac1r}, & r\neq\frac12 \\ \displaystyle
    1+\chi^2 \ln \chi^2 + C\chi^2,  & r=\frac12
\end{cases}
\ee
where $C$ is an integration constant that has an upper limit, $C\leq C_*$, with $C_*=2r^{1+\frac{1}{2r}}/(1-2r)$ for $r\neq1/2$, and $C_* = \ln{2} -2 $ for $r=1/2$. In Fig.~\ref{figorbit}, we display these orbits for $r=1/8,1/4,1/2$ and $7/8$, with several values of $C$. Note that the curves associated to the orbits are different for each value of $r$.

With the goal of finding novel results in the model in which the fields $\phi$ and $\chi$ are coupled to $\psi$ via $f(\psi)$ and $g(\psi)$, let us review the standard BNRT model, in which $f(\psi)=g(\psi)=1$, whose solutions will be useful. In this case, the first-order equations \eqref{foblochfg} take the form
\bes\label{fobloch}
\bal
\phi' &= 1 - \phi^2 - r\chi^2, \\
\chi' &= -2 r \phi \chi.
\eal\ees
The above system of differential equations can be decoupled by using the orbit equation \eqref{bnrtorbitr}. One may show that, by solving
\be
\chi' = \begin{cases}\displaystyle
            \mp 2r \chi \sqrt{1-\frac{r}{1-2r}\chi^2 + C |\chi|^{\frac1r}}, & r\neq\frac12 \\ \displaystyle
            \mp 2r\chi \sqrt{1+\chi^2 \ln \chi^2 + C\chi^2},  & r=\frac12,
            \end{cases} 
\ee
the solution $\phi(x)$ can be obtained from the orbit in Eq.~\eqref{bnrtorbitr}. The solutions of this equation were investigated in Ref.~\cite{izquierdoorbit}, where it was shown that one cannot calculate analytical solutions for all $r$ and $C$. There are two known possibilities to determine analytical solutions for a general $r$. The first one is to consider $C\to-\infty$ to get a straight-line orbit that connects the minima $v_{h\pm}$, leading to the solution $\phi(x) = \tanh(x)$ and $\chi(x)=0$. The second possibility is to take $C=0$, which leads to an elliptic orbit that also connects $v_{h\pm}$. The solution in this situation is $\phi(x) = \tanh(2rx)$ and $\chi(x) = \pm\sqrt{\frac{1-2r}{r}}\,\sech(2rx)$, with the restriction $0<r<1/2$, where the upper/lower sign in $\chi(x)$ stands for the curves in the orbits above/below the $\phi$-axis. These solutions, associated to the straight-line and elliptic orbits, are only valid for a single value of $C$ and several values of $r$. One may obtain $C$-dependent analytical solutions for a single value of $r$ by fixing $r=1/4$, which makes the maximum value of $C$ become $C_*=1/16$. For $C<1/16$, it has the form
\bes\label{solc}
\bal
\phi(x) &= \frac{c\sinh(x)}{1 + c\cosh(x)},\\
\chi(x) &= \pm\frac{2}{\sqrt{1+ c\cosh(x)}},
\eal
\ees
where $c=\sqrt{1-16\,C}$, such that $c$ ranges from $0$ to $\infty$, with the energy density 
\be\label{rhoc}
\rho_1(x)=\frac{c^2(c+\cosh(x))^2}{(1+c\cosh(x))^4} + \frac{c^2\sinh^2(x)}{(1+c\cosh(x))^3}.
\ee
The set of solutions \eqref{solc} connects the minima $v_{h\pm}$.

The critical case, in which $C=C_*=1/16$, leads to solutions connecting the neighbor minima of the diamond. The solution connecting $v_{v+}$ to $v_{h+}$ is
\bes\label{sol116}
\bal
\phi_{v+ h+}(x) = \frac{1}{2}+\frac{1}{2}\tanh\left(\frac{x}{2}\right),\\
\chi_{v+ h+}(x) = \sqrt{2-2\tanh\left(\frac{x}{2}\right)}.
\eal
\ees
For the minima $v_{v-}$ to $v_{h+}$, one can show that the solution is
\bes\label{sol116v-h+}
\bal
\phi_{v-h+}(x) = \frac{1}{2}+\frac{1}{2}\tanh\left(\frac{x}{2}\right),\\
\chi_{v-h+}(x) = -\sqrt{2-2\tanh\left(\frac{x}{2}\right)}.
\eal
\ees
Both the solutions in Eqs.~\eqref{sol116} and \eqref{sol116v-h+} have energy density given by
\be\label{rho116}
\rho_1(x)=\frac{e^x+2}{4\,(1+\cosh(x))^2}.
\ee
The minima $v_{h-}$ and $v_{v-}$ are linked by the solution 
\bes\label{sol116h-v-}
\bal
\phi_{h-v-}(x) = -\frac{1}{2}+\frac{1}{2}\tanh\left(\frac{x}{2}\right),\\
\chi_{h- v-}(x) = -\sqrt{2+2\tanh\left(\frac{x}{2}\right)}.
\eal
\ees
The last solution for the critical value of $C$ is
\bes\label{sol116h-v+}
\bal
\phi_{h-v+}(x) = -\frac{1}{2}+\frac{1}{2}\tanh\left(\frac{x}{2}\right),\\
\chi_{h-v+}(x) = \sqrt{2+2\tanh\left(\frac{x}{2}\right)}.
\eal
\ees
It connects $v_{h-}$ to $v_{v+}$. The solutions in Eq.~\eqref{sol116h-v-} and \eqref{sol116h-v+} have the energy density in the form
\be\label{rho116-}
\rho_1(x)=\frac{e^{-x}+2}{4\,(1+\cosh(x))^2}.
\ee
The energy densities in Eqs.~\eqref{rho116} and \eqref{rho116-} differ from each other, but can be related through the change $x\to-x$, revealing an asymmetry in the structure.

\subsection{The case $f(\psi)=g(\psi)$}\label{secf}
Let us first illustrate the case in which $f(\psi)=g(\psi)$. In this situation, we can introduce a novel coordinate, $\xi$, defined by
\be\label{xif}
\frac{d\xi}{dx} = \frac{1}{f(\psi(x))}\implies \xi = \int \frac{dx}{f(\psi(x))}.
\ee
From this, we see that $\xi$ depends explicitly on the source field as it feeds the mediator function $f(\psi(x))$. The above expression helps us to write the first-order equations \eqref{foblochfg} as 
\bes\label{foblochxi}
\bal
\phi_\xi &= 1 - \phi^2 - r\chi^2, \\
\chi_\xi &= -2 r \phi \chi,
\eal\ees
where the subscripts represent the derivative with respect to $\xi$. The above equations lead to solutions with argument $\xi$, which can be used to induce geometric modifications in the structures \cite{constbloch,const1,const2,const3}. Interestingly, only in the case $f(\psi)=g(\psi)$ the model supports the same orbit in Eq.~\eqref{bnrtorbitr}, and solutions in Eqs.~\eqref{solc}, \eqref{sol116}, \eqref{sol116v-h+}, \eqref{sol116h-v-} and \eqref{sol116h-v+} with the change $x\to\xi$. The explicit expression for $\xi$ must be obtained from Eq.~\eqref{xif}. We remark, however, that the energy density associated to the aforementioned solutions are given, respectively, by
\be\label{rhocxif}
\begin{aligned}
    \rho_1(x)&=\frac{c^2(c+\cosh(\xi(x)))^2}{f(\psi(x))\,(1+c\cosh(\xi(x)))^4} \\
    &+ \frac{c^2\sinh^2(\xi(x))}{f(\psi(x))\,(1+c\cosh(\xi(x)))^3}.
\end{aligned}
\ee
for Eq.~\eqref{rhoc}, and
\be\label{rho116xif}
\rho_1(x)=\frac{(e^{\pm\xi(x)}+2)}{4f(\psi(x))\,(1+\cosh(\xi(x)))^2},
\ee
in which the upper and lower signs stands for the energy density corresponding to \eqref{rho116} and \eqref{rho116-}, respectively. Notice that $f(\psi)$ is explicitly present in the energy density. Therefore, it can be used to induce modifications in specific points of the core and/or the tail of the structure. The expressions in Eqs.~\eqref{rhoc}, \eqref{rho116} and \eqref{rho116-} are recovered for $f(\psi)=1$, as expected.

To determine the form of $\xi(x)$, we must provide the function $f(\psi)$. Next, we present some possibilities with analytical results. We first consider
\be\label{fcos}
f(\psi)=\frac{1}{\cos^2(n\pi\psi)},
\ee
where $n$ is an integer number. One must keep in mind that, following our first-order formalism, the field $\psi$ that appears in the above expression is the solution of Eq.~\eqref{fopsi}. Since we are considering $\psi$ to be driven by Eq.~\eqref{qpsi4}, we have $\psi(x) = \tanh(\alpha x)$. So, the above function becomes $f(\psi(x)) = 1/\cos^2(n\pi\tanh(\alpha x))$, which engender $2n$ symmetric divergent points outside the origin. It was shown in Ref.~\cite{constbloch} that this function leads to plateaus in the solutions of the elliptic orbit of the BNRT model and valleys in their respective energy densities. From Eq.~\eqref{xif}, one gets
\be\label{xici}
\xi(x) = \frac{x}{2} + \frac{1}{4\alpha}\left(\text{Ci}(\eta_+)-\text{Ci}(\eta_-)\right),
\ee
where $\eta_\pm = 2n\pi(1\pm\tanh(\alpha x))$ and $\text{Ci}(z)$ is the cosine integral function \cite{const1}. This expression defines the coordinate which replaces $x$ in the solutions \eqref{solc}, \eqref{sol116}, \eqref{sol116v-h+}, \eqref{sol116h-v-} and \eqref{sol116h-v+}, whose energy densities now must be calculated from Eq.~\eqref{rhophichi}, having the form in Eqs.~\eqref{rhocxif} and \eqref{rho116xif}, with $f(\psi)$ as in Eq.~\eqref{fcos}.

Another possibility is to consider a function $f(\psi)$ which includes a divergent point at the center of $\psi(x)$. To do so, we take
\be\label{fsin}
f(\psi)=\frac{1}{\sin^2\left(\left(n+\frac12\right)\pi\psi\right)},
\ee
such that $f(\psi(x))=1/\sin^2\!\left(\left(n+1/2\right)\pi\tanh(\alpha x)\right)$ presents $2n$ symmetric divergent points outside the origin and one divergent point located at the origin. In this situation, the geometrical coordinate \eqref{xif} that substitutes $x$ in the solutions ($\phi(x),\chi(x)$) takes the same form of Eq.~\eqref{xici}, with $\eta_\pm = (2n+1)\pi(1\pm\tanh(\alpha x))$. The energy densities \eqref{rhocxif} and \eqref{rho116xif} must now be taken with $f(\psi)$ written as in the above expression. 

The functions in Eqs.~\eqref{fcos} and \eqref{fsin} lead to modifications in the core of the structure, making the solutions attain plateaus and their respective energy densities get valleys whose quantity is controlled by the parameter $n$. We now use $f(\psi)$ to compactify the structure, shrinking its tail to a compact space. Inspired by the recent work \cite{const3}, where we have used a similar procedure to get compact kinks, we consider
\be\label{fcomp}
f(\psi) = (1-3\psi^2)^2,
\ee
where the factor $3$ at the $\psi^2$ term was chosen for convenience, without loss of generality. One may replace it for any real number greater than $1$. Since we are taking $\psi(x)=\tanh(\alpha x)$, the above expression becomes $f(\psi(x)) = (1-3\tanh^2(\alpha x))^2$. It vanishes at two symmetric points around the origin, which is required to compactify the structure. In this situation, the geometrical coordinate $\xi$ from Eq.~\eqref{xif} that replaces $x$ in the solutions becomes
\be
\xi(x) = \begin{cases}
\displaystyle\frac{x}{4} - \frac{3\tanh(\alpha x)}{4\alpha (3\tanh^2(\alpha x) -1)} -c,\,\,&|x|\leq x_c\\
\textrm{sgn}(x) \,\infty, \,\, & |x|>x_c
\end{cases},
\ee
where $x_c=\text{arctanh}(1/\sqrt{3})/\alpha$ delimits the compact space. It shrinks the solutions to the width $2x_c$. The energy densities can be calculated from Eqs.~\eqref{rhocxif} and \eqref{rho116xif} with $f(\psi)$ substituted by the expression in \eqref{fcomp}. Near the points of compactification, $x\approx\pm x_c$, one gets 
\be\label{xi*} \xi(x)\approx \xi_\pm^*\equiv-\frac{3}{16\alpha^2(x\mp x_c)},
\ee 
for $x$ approaching $\pm x_c$ through points inside the compact space. This makes the solution in Eq.~\eqref{solc} with the change $x\to\xi$ behave as $\phi(x)\approx1-2e^{-\xi^*_+}/c$ and $\chi(x) \approx 2\sqrt{2/c}\,e^{-\xi^*_+/2}$ for $x\to x_c$, and $\phi(x)\approx-1+2e^{\xi^*_-}/c$ and $\chi(x) \approx 2\sqrt{2/c}\,e^{\xi^*_-/2}$ for $x\to-x_c$. So, the falloff of the solutions is exponential in $\xi^*$, which depends on $x$ as in \eqref{xi*}, near the points of compactification. A similar analysis can be done for the solution \eqref{sol116}, which connects $v_{v+}$ to $v_{h+}$, such that we get $\phi(x)\approx 1- e^{-\xi^*_+}$ and $\chi(x)\approx 2 e^{-\xi^*_+/2}$ for $x\to x_c$, and $\phi(x)\approx e^{\xi^*_-}$ and $\chi(x)\approx2-e^{\xi^*_-}$ for $x\to-x_c$. The behavior of the solutions \eqref{sol116v-h+}, \eqref{sol116h-v-} and \eqref{sol116h-v+} is similar, so we omit it here.

\begin{figure}[t!]
    \centering
    \includegraphics[width=4.2cm]{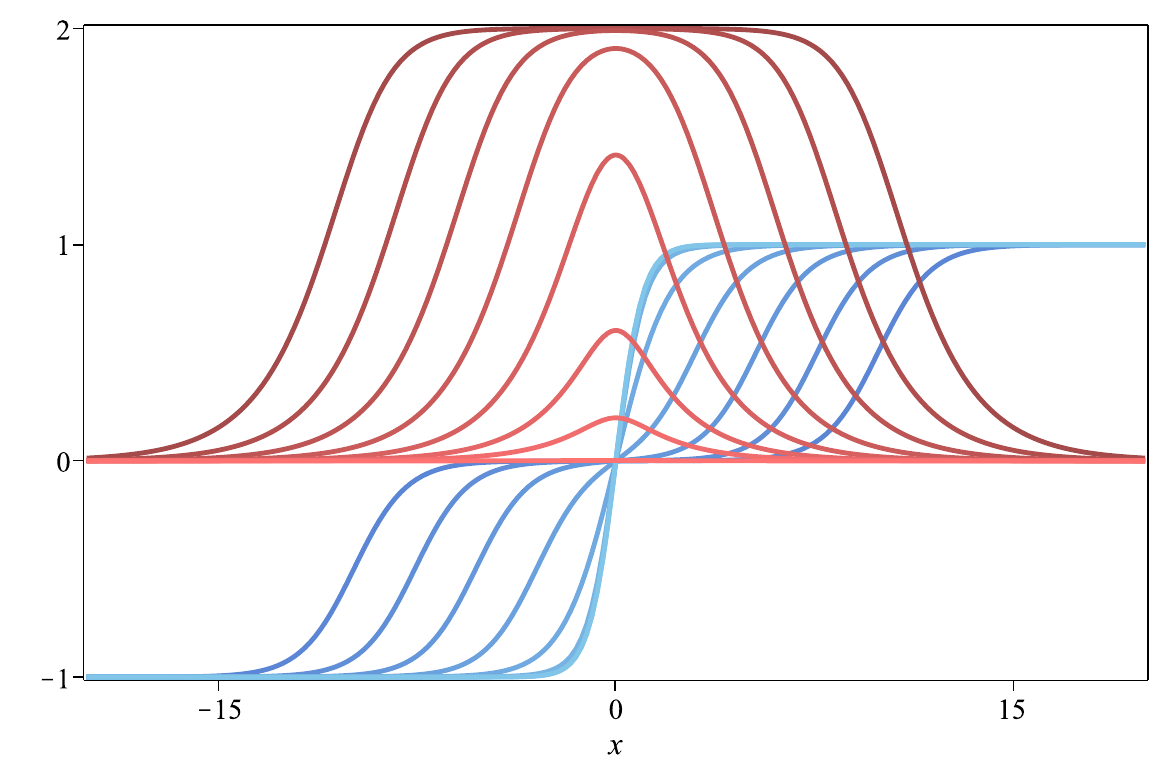}
    \includegraphics[width=4.2cm]{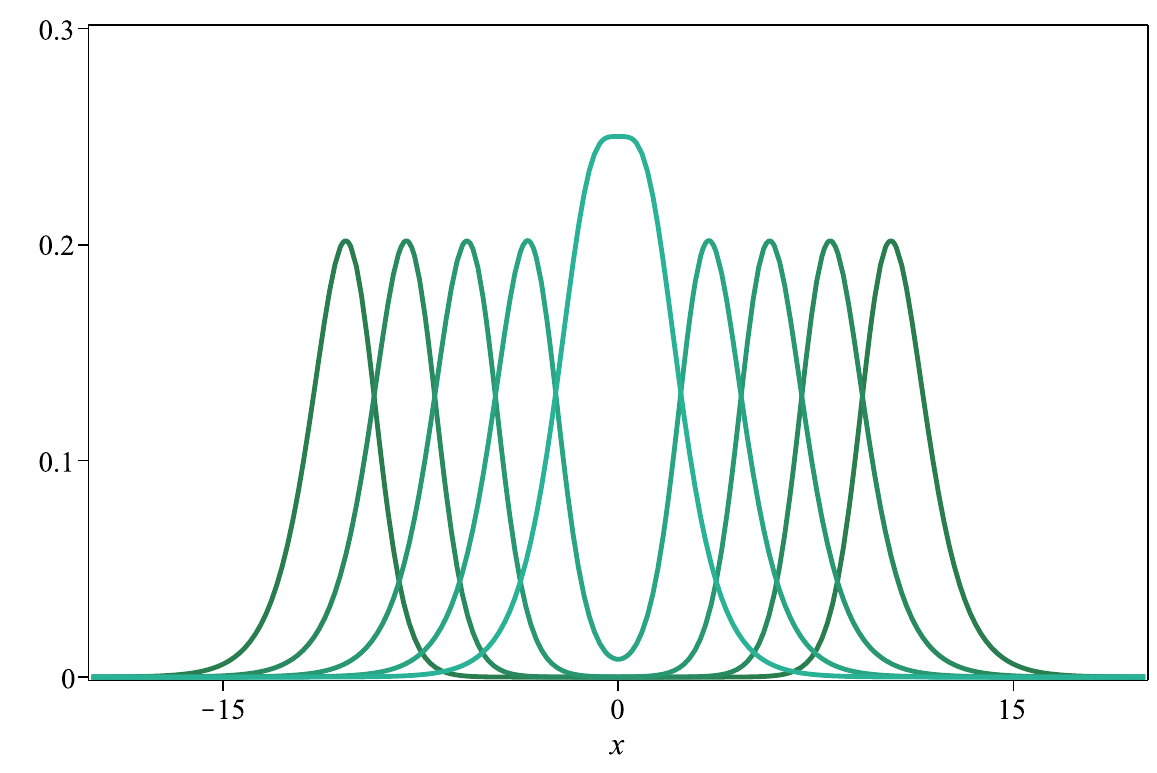}
    \includegraphics[width=4.2cm]{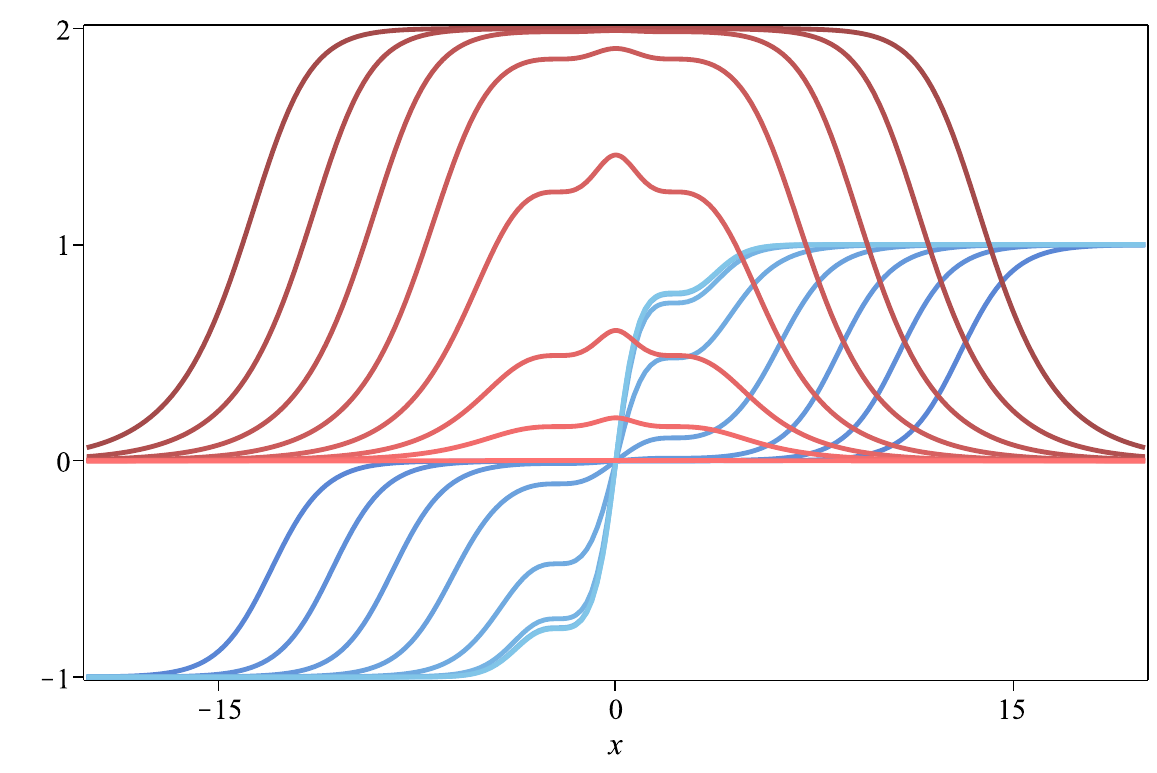}
    \includegraphics[width=4.2cm]{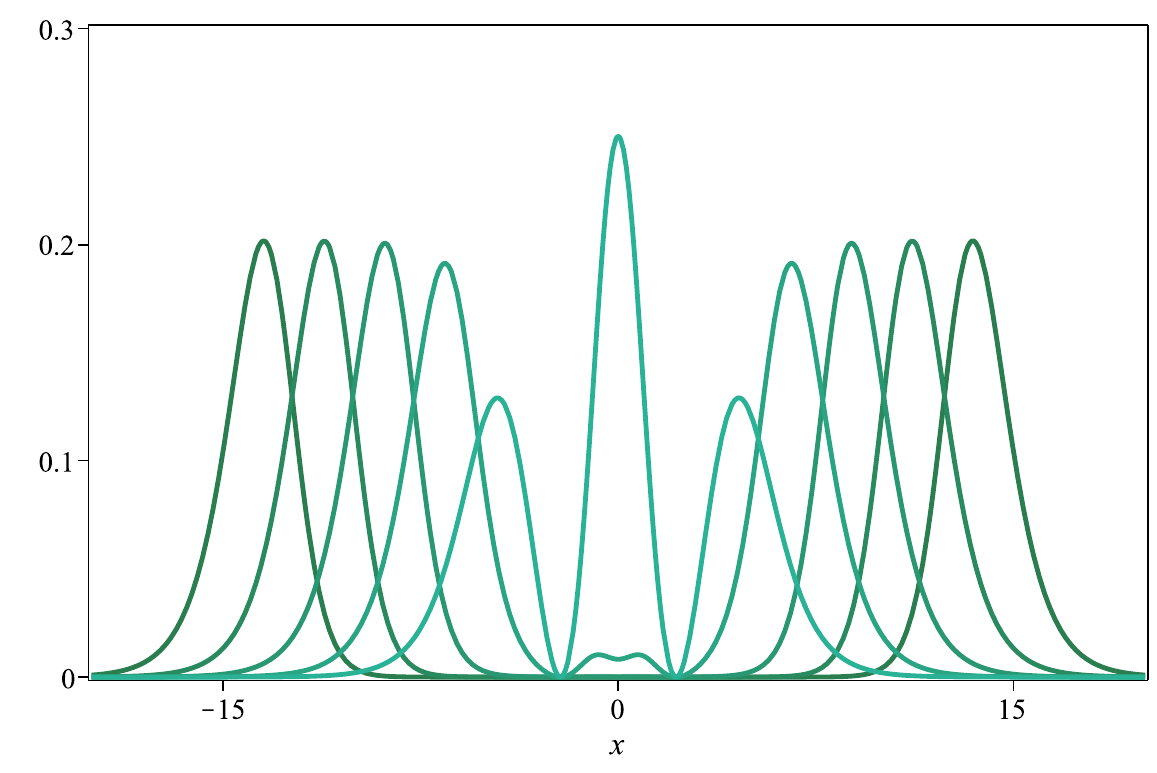}
    \includegraphics[width=4.2cm]{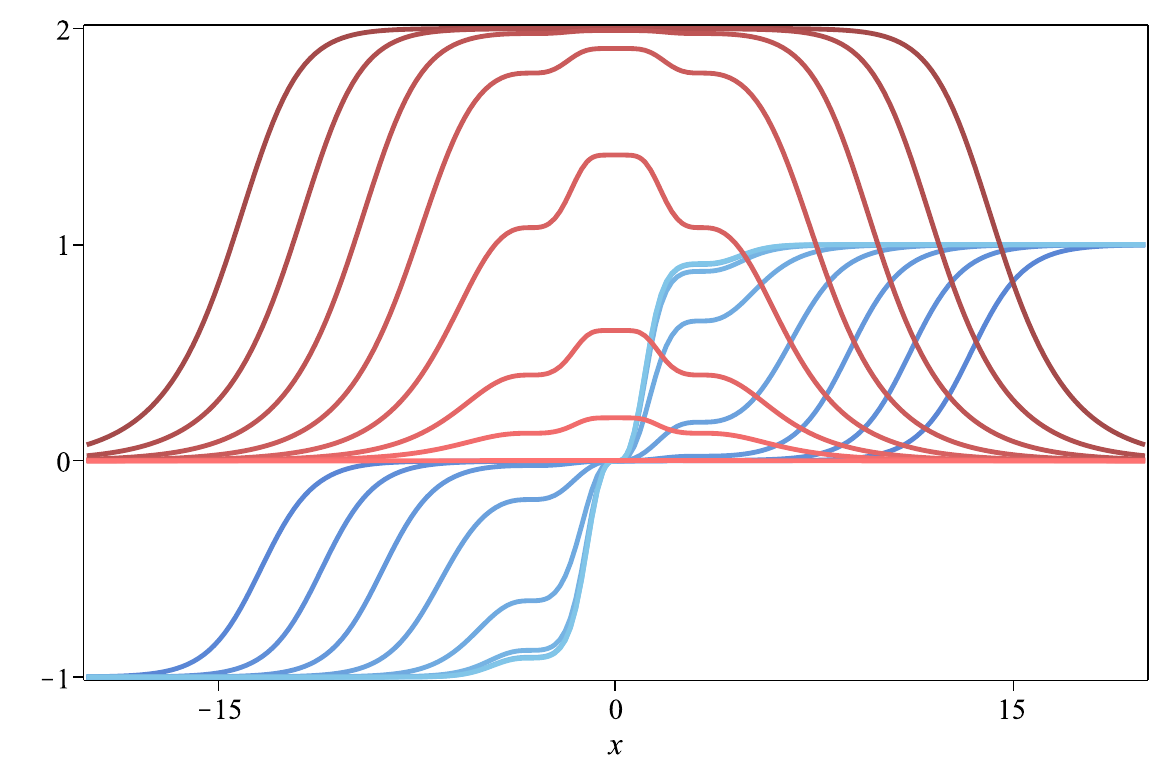}
    \includegraphics[width=4.2cm]{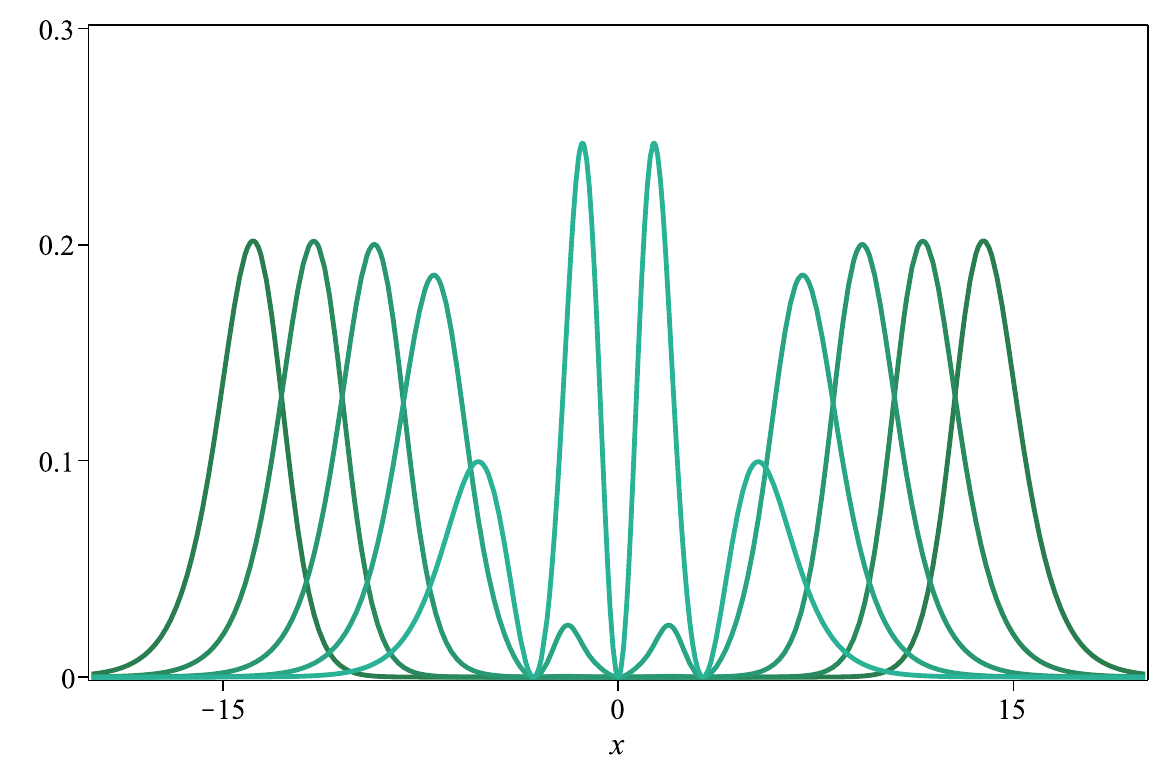}
    \includegraphics[width=4.2cm]{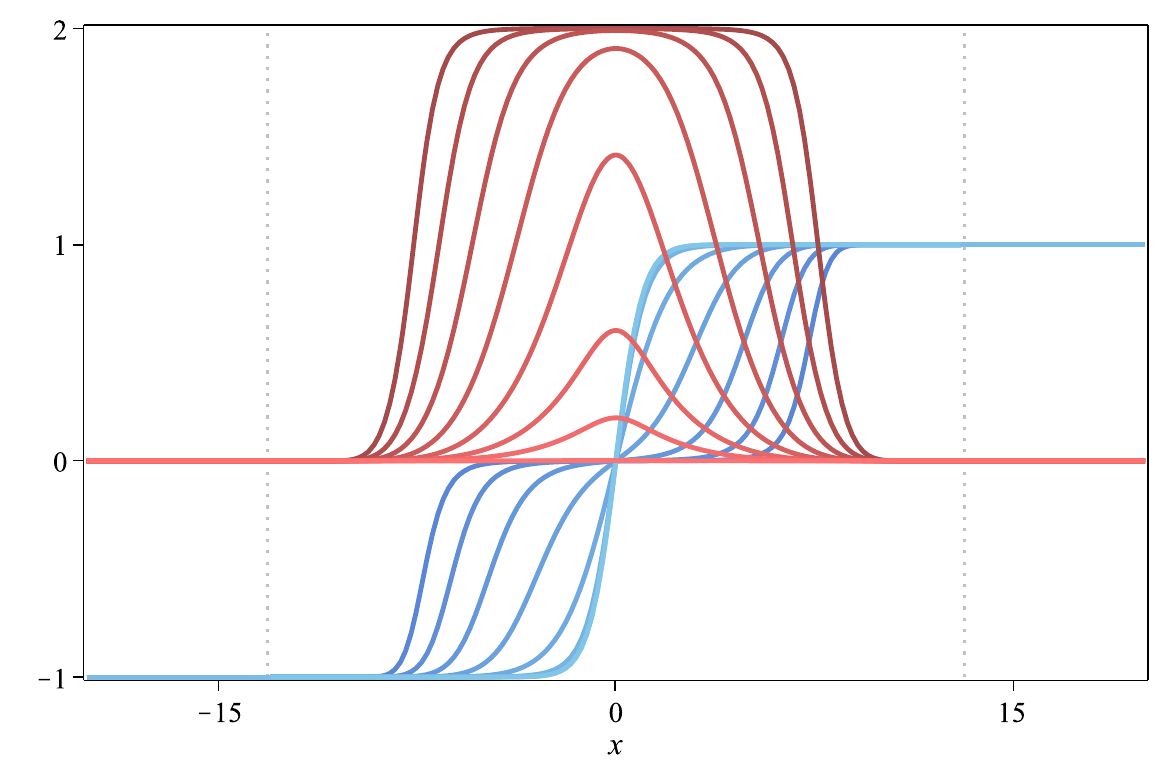}
    \includegraphics[width=4.2cm]{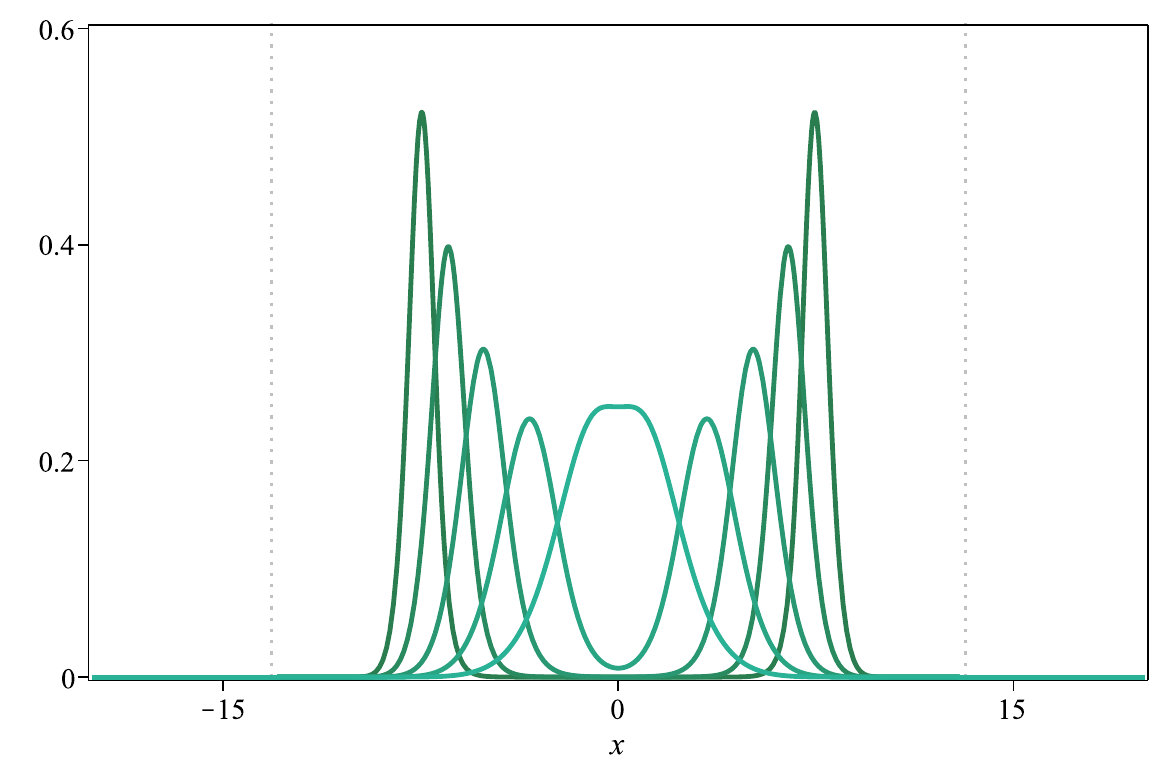}
    \caption{In the left panels, we display the solutions $\phi (x)$ (blue) and $\chi(x)$ (red) in Eq.~\eqref{solc} with the change $x\to\xi$, in the case $f(\psi)=g(\psi)$ for $r=1/4$ and $c= 10^k$, with $k=-4,-3,\ldots 2$ and the limit $c\to\infty$. In the right panels, we show the associated energy density $\rho_1$ (green) for $k=-4,-3,\ldots 0$, with the positive values of $k$ excluded for the sake of clarity, as they have a similar profile of the case $k=0$, but their peak is too tall. The top, middle-top, middle-bottom and bottom panels represent the case in which $f(\psi)=1$, $f(\psi)=1/\cos^2(n\pi\psi)$ in Eq.~\eqref{fcos} with $n=1$ and $\alpha=1/4$, $f(\psi)=1/\sin^2((n+1/2)\pi\psi)$ in Eq.~\eqref{fsin} with $n=1$ and $\alpha=1/4$, and $f(\psi) = (1-3\psi^2)^2$ in Eq.~\eqref{fcomp} with $\alpha=1/20$, respectively. The dotted vertical lines in gray delimit the compact space $[-x_c,x_c]$. The colors get lighter as $k$ increases.}
    \label{figparsolrho1}
\end{figure}

To illustrate how the functions discussed above modify the structure, in Fig.~\ref{figparsolrho1} we display the solutions \eqref{solc} with $x\to\xi$ and the respective energy density \eqref{rhocxif} for the standard case ($f(\psi)=1$) and the other functions in Eqs.~\eqref{fcos}, \eqref{fsin} and \eqref{fcomp}. At it is well known, in the standard case, the slope of the solutions at $x=0$ is controlled by $c$, which leads to the formation of a plateau in $\phi(x)$ and to the enlargement of the plateau in $\chi(x)$ as $c$ approaches zero. This gives rise to a splitting in the energy density, such that the separation of the lump-like parts gets larger as $c$ decreases; see the top panels. The model with the function \eqref{fcos} presents the formation of $2n$ symmetric plateaus outside the center of the solutions for all $c$, leading to an energy density with several lump-like portions; see the middle-top panels. By considering the function \eqref{fsin}, one also gets the inclusion of a plateau with null slope in the center of the solutions for all $c$, which creates a hole in the energy density at this point; see the middle-bottom panels. The bottom panels display the solutions and energy density associated to the compact structure induced by the function \eqref{fcomp}. As far as we know, this is the first time that we have seen compact solutions in multifield models. Also, the functions considered above can be used in any solution of the BNRT model \eqref{pbnrt}, which allows to induce modifications in non-analytical solutions for all values of $r$, which can only be obtained from numeric methods.

\begin{figure}[t!]
    \centering
    \includegraphics[width=4.2cm]{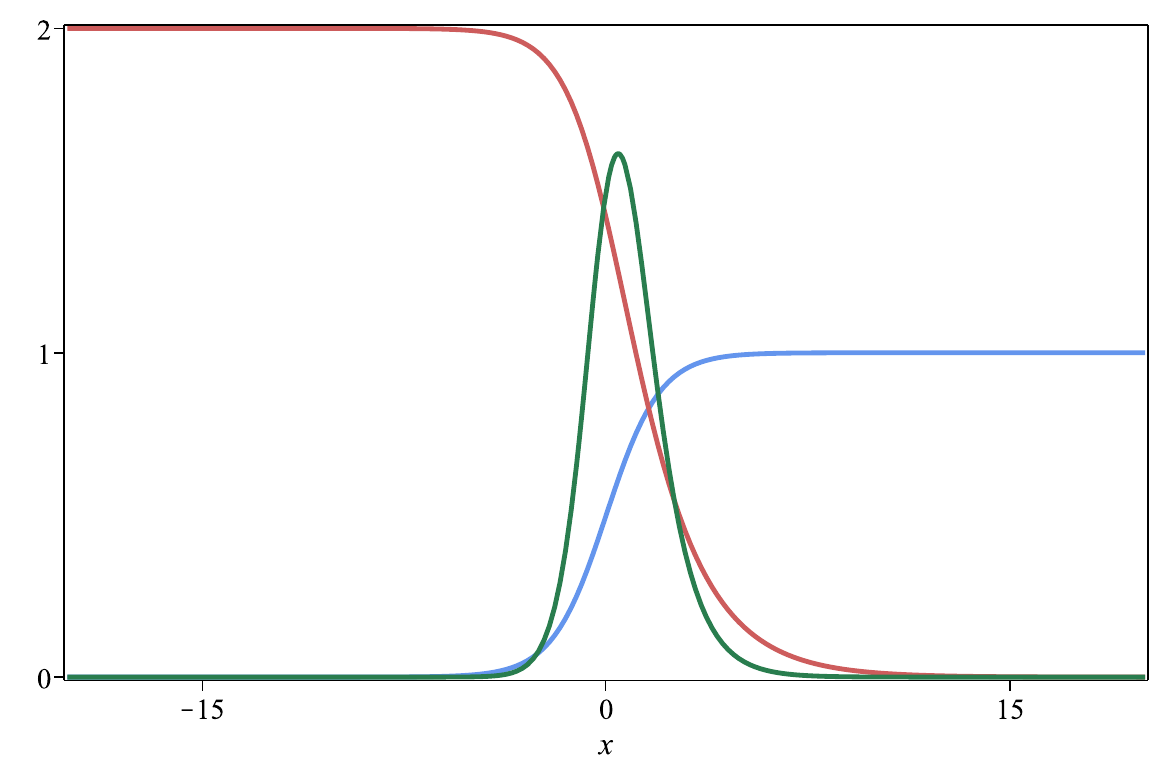}
    \includegraphics[width=4.2cm]{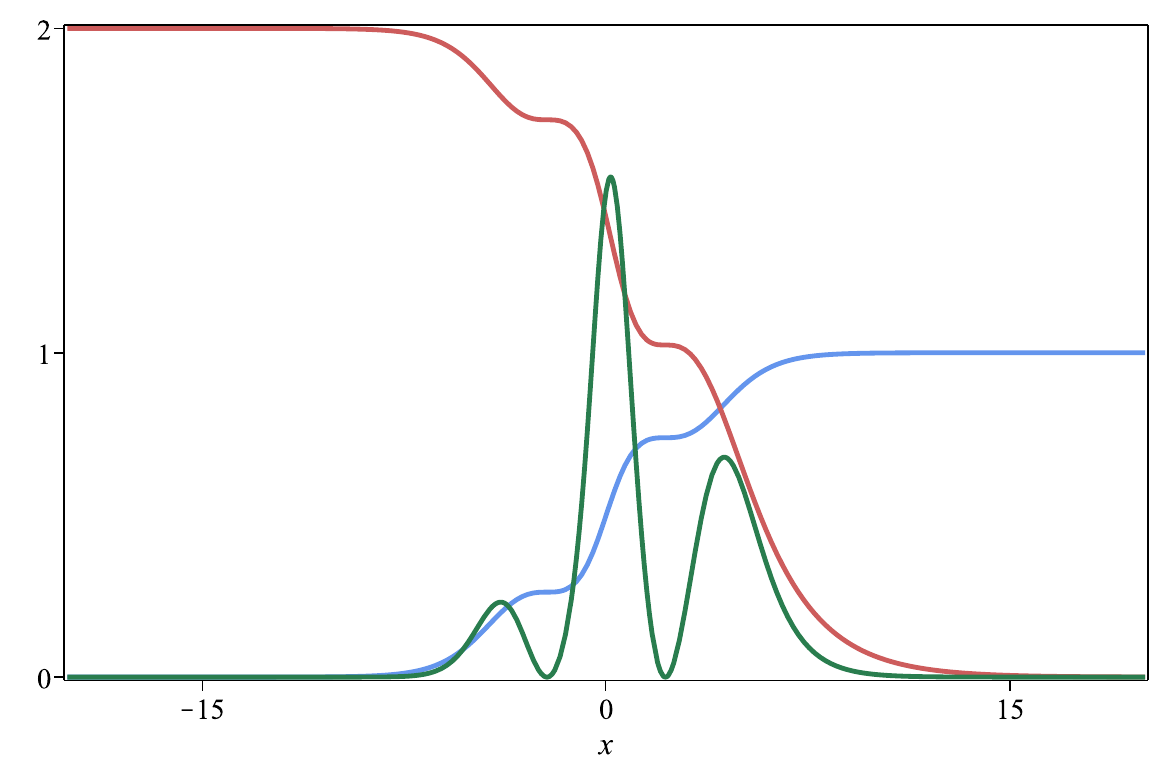}
    \includegraphics[width=4.2cm]{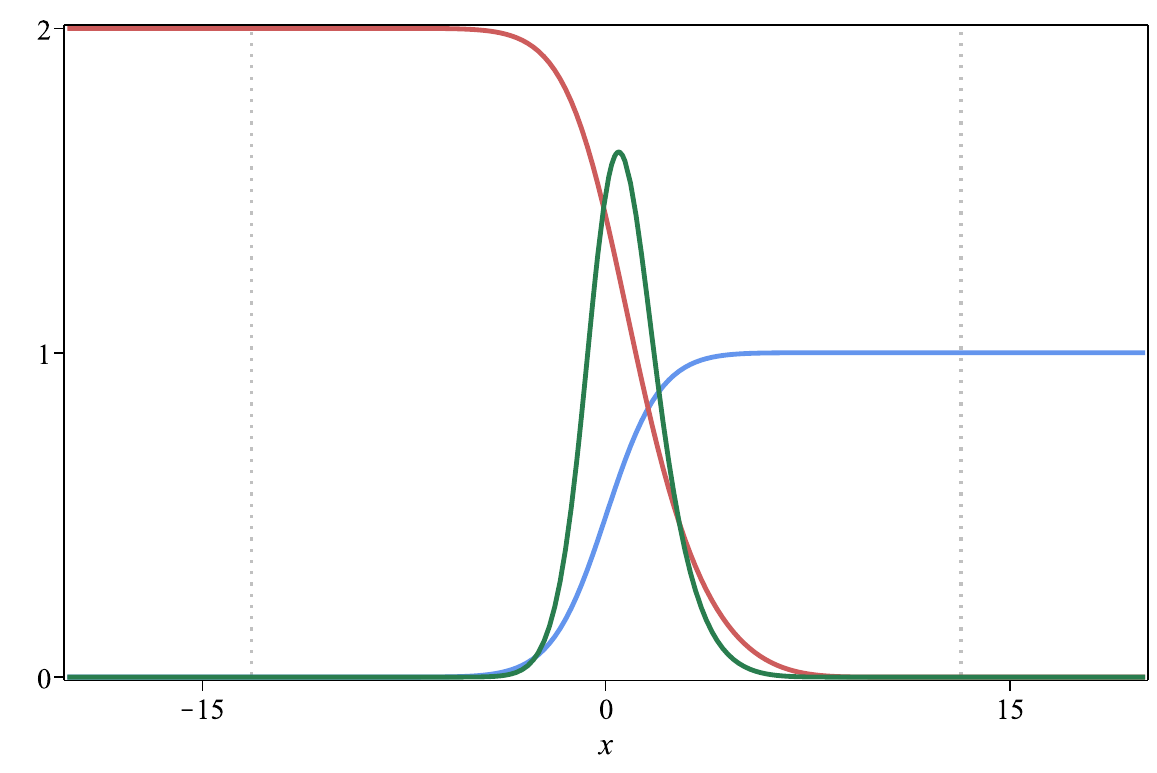}
    \includegraphics[width=4.2cm]{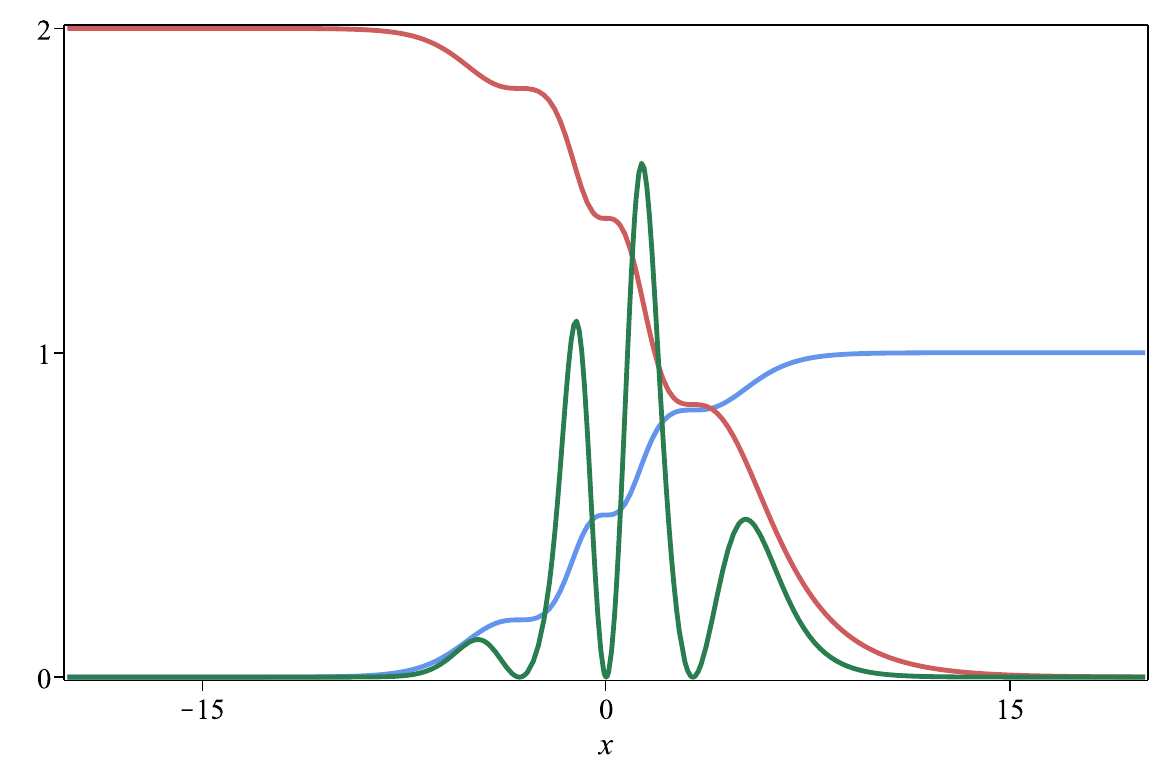}
    \caption{The solution $\phi (x)$ (blue) and $\chi(x)$ (red) in Eq.~\eqref{sol116} with the change $x\to\xi$ and the associated energy density $8\rho_1$ (green) for the models in which $f(\psi)=g(\psi)$, with $f(\psi)=1$ (top left), $f(\psi)=1/\cos^2(n\pi\psi)$ with $n=1$ and $\alpha=1/4$ (top right), $f(\psi) = (1-3\psi^2)^2$ with $\alpha=1/8$ (bottom left) and $f(\psi)=1/\sin^2((n+1/2)\pi\psi)$ with $n=1$ and $\alpha=1/4$ (bottom right). The dotted vertical lines in gray delimit the compact space.}
    \label{figparsolrho2}
\end{figure}

In Fig.~\ref{figparsolrho2}, we display the solution in Eq.~\eqref{sol116} with the change $x\to\xi$ and the energy density in Eq.~\eqref{rho116xif} for the standard case ($f(\psi)=1$) and the functions in Eqs.~\eqref{fcos}, \eqref{fsin} and \eqref{fcomp}. Notice that, in the standard case depicted in the top-left panel, both $\phi(x)$ and $\chi(x)$ has an asymmetric kink-like profile, each one connecting two distinct values asymptotically; their associated energy density engender an asymmetric lump-like shape without the splitting that appears in the solution discussed in the previous paragraph. For the function in Eq.~\eqref{fcos} plotted in the top-right panel, we see that both $\phi(x)$ and $\chi(x)$ attain plateaus at symmetric points outside the origin such that the energy density gets split into $2n+1$ lumps. In the bottom-right panel, we display the model with the function \eqref{fsin}, with includes the modification in the origin, leading to $2n+1$ plateaus in the solution and $2n+2$ lumps in the energy density. The bottom-left panel presents the compact structure generated via the function \eqref{fcomp}.

\subsection{The case $f(\psi)\neq g(\psi)$}
We now turn our attention to the case in which the functions that modify the dynamical terms in the Lagrangian density \eqref{lmodel1}, $f(\psi)$ and $g(\psi)$, are not equal. This situation is much more complicated and requires the use of numerical approach. In contrast to the case discussed in Sec.~\ref{secf}, one cannot solve Eq.~\eqref{orbiteq} to get the orbit without knowing the solutions explicitly. The orbit must be constructed \emph{after} calculating the solutions. Moreover, in this case, one cannot define the geometrical coordinate $\xi$ to write the solutions in terms of the ones associated to the standard case ($f(\psi)=g(\psi)=1$). This means that Eq.~\eqref{foblochfg} must be particularly solved for each pair $f(\psi)$ and $g(\psi)$. To feed these mediator functions, we consider the source field to be described by Eq.~\eqref{qpsi4}, so one has $\psi(x)=\tanh(\alpha x)$. The other fields are described by the BNRT model \eqref{pbnrt}, obeying the first-order equations
\be\label{firstfg}
\phi' = \frac{1 - \phi^2 - r\chi^2}{f(\tanh(\alpha x))} \quad\text{and}\quad \chi' = -\frac{2 r \phi \chi}{g(\tanh(\alpha x))}.
\ee

\begin{figure}[t!]
    \centering
    \includegraphics[width=4.2cm]{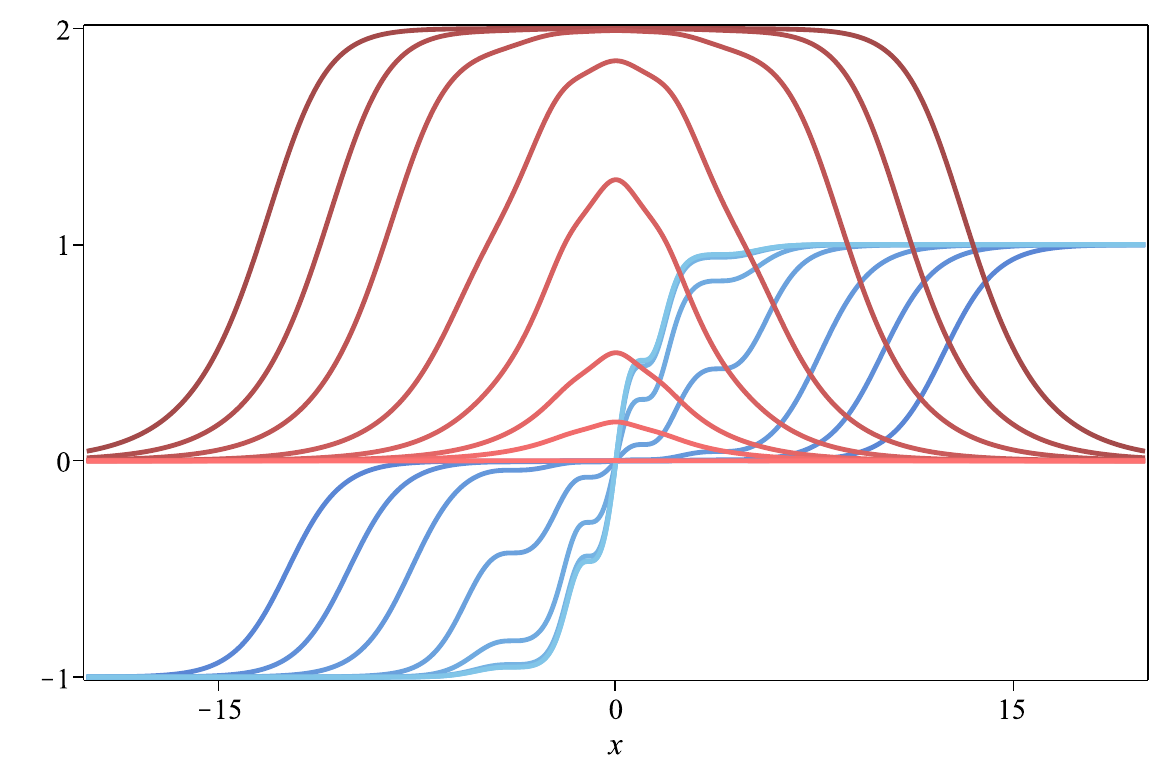}
    \includegraphics[width=4.2cm]{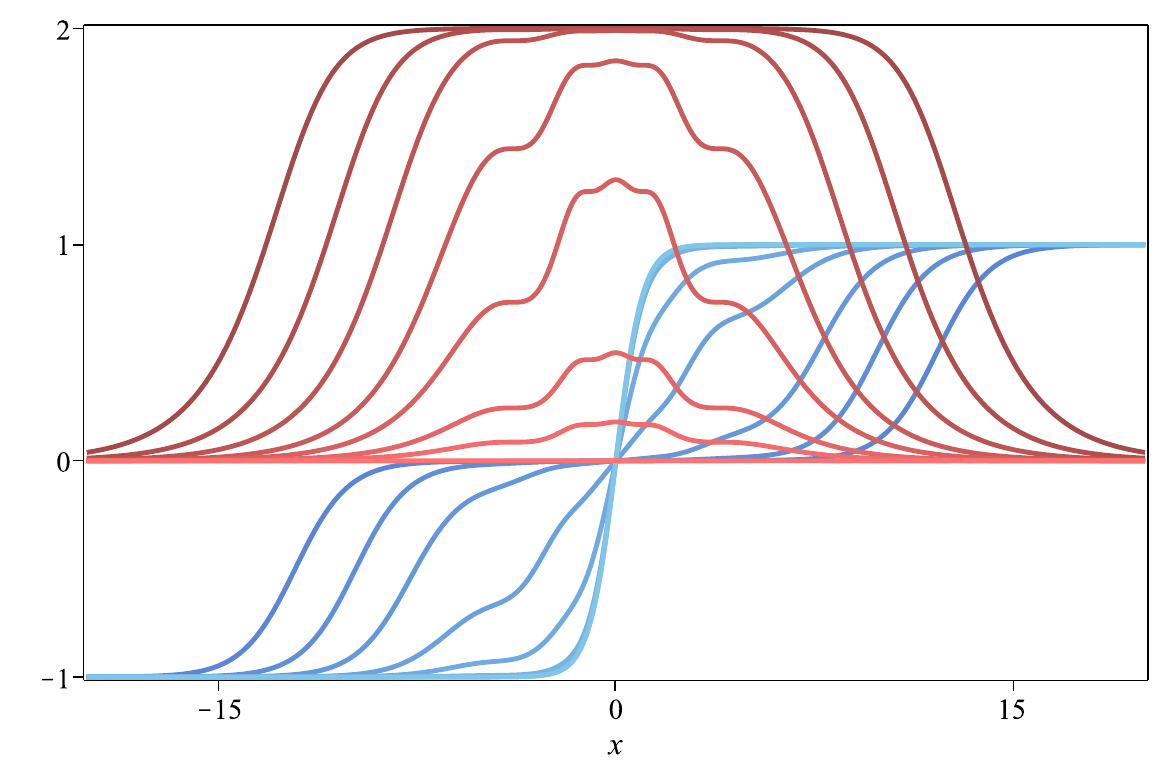}
    \includegraphics[width=4.2cm]{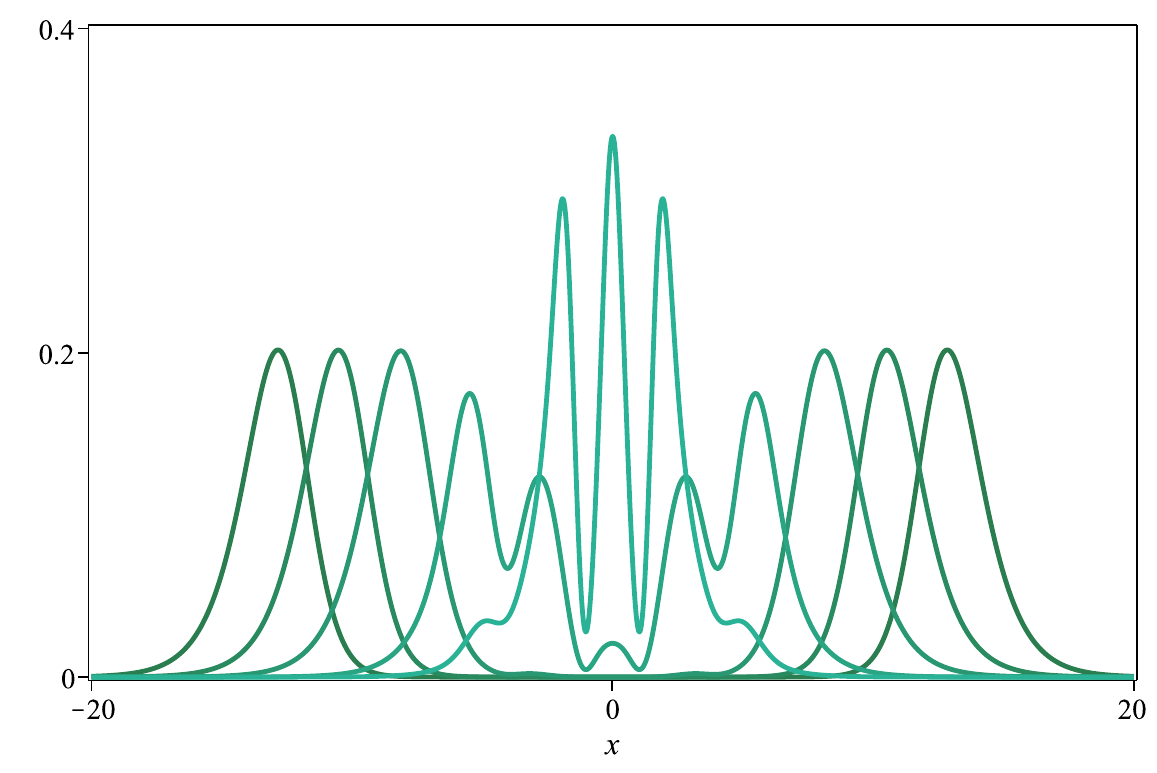}
    \includegraphics[width=4.2cm]{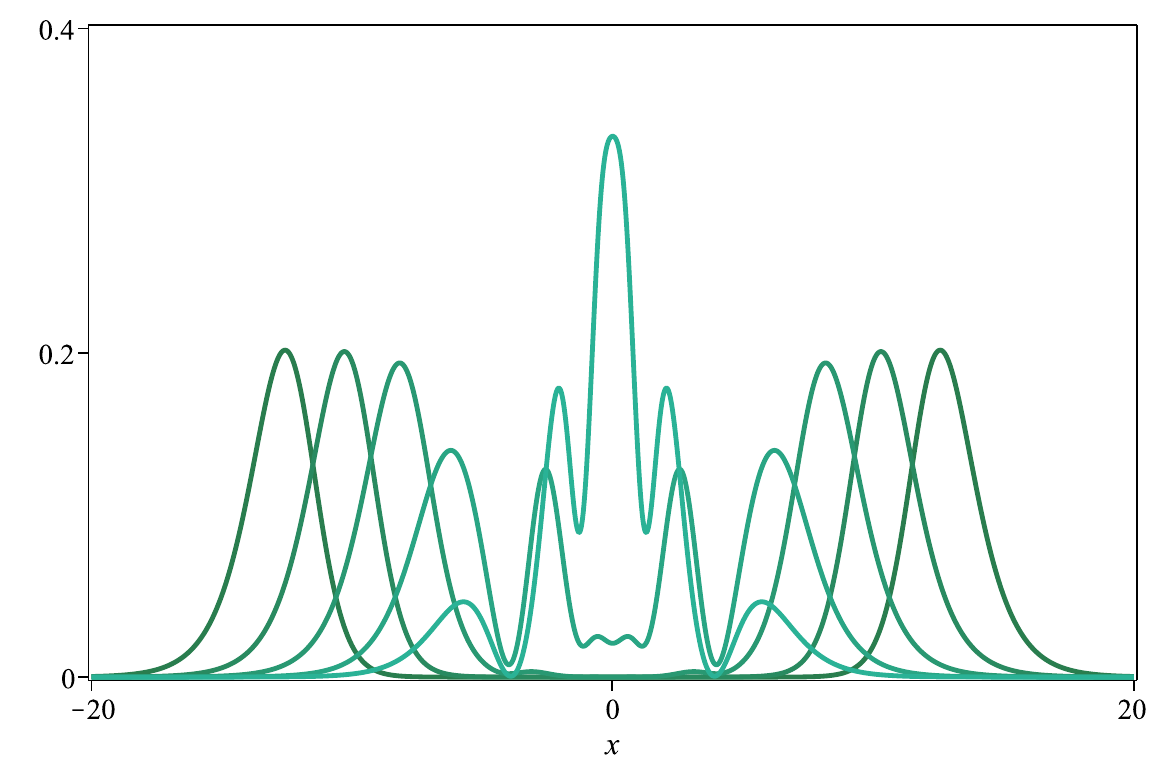}
    \includegraphics[width=4.2cm]{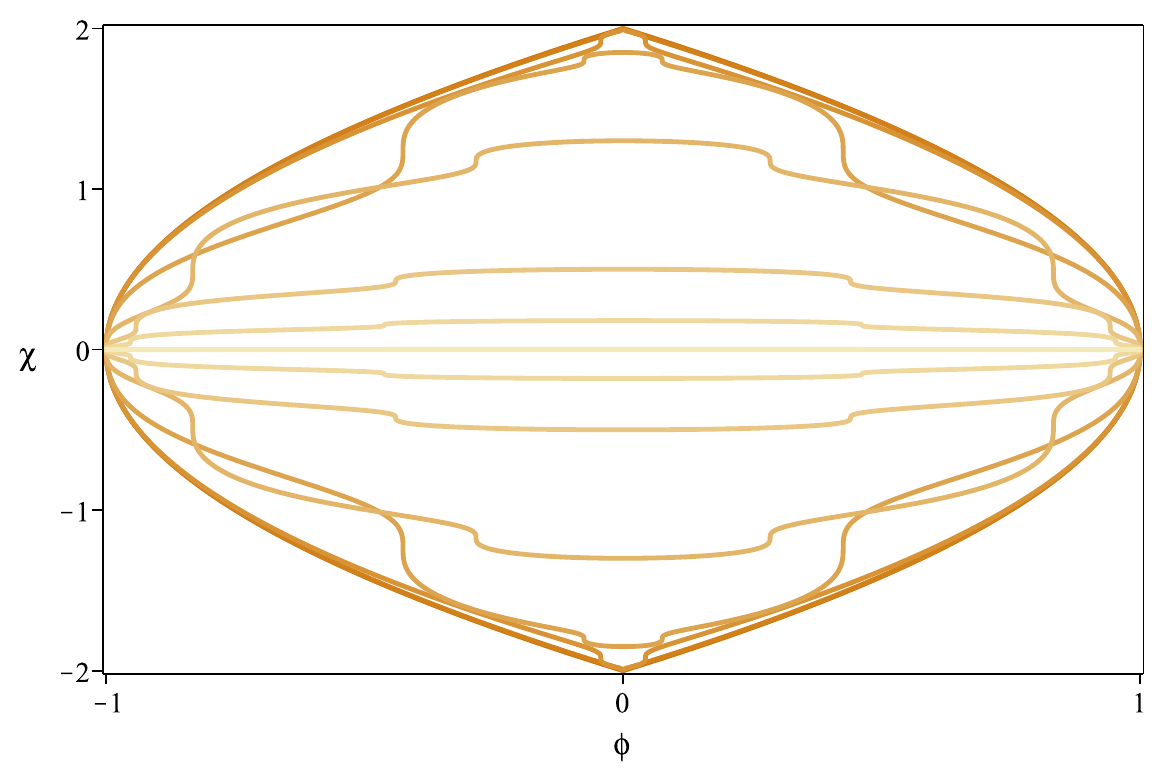}
    \includegraphics[width=4.2cm]{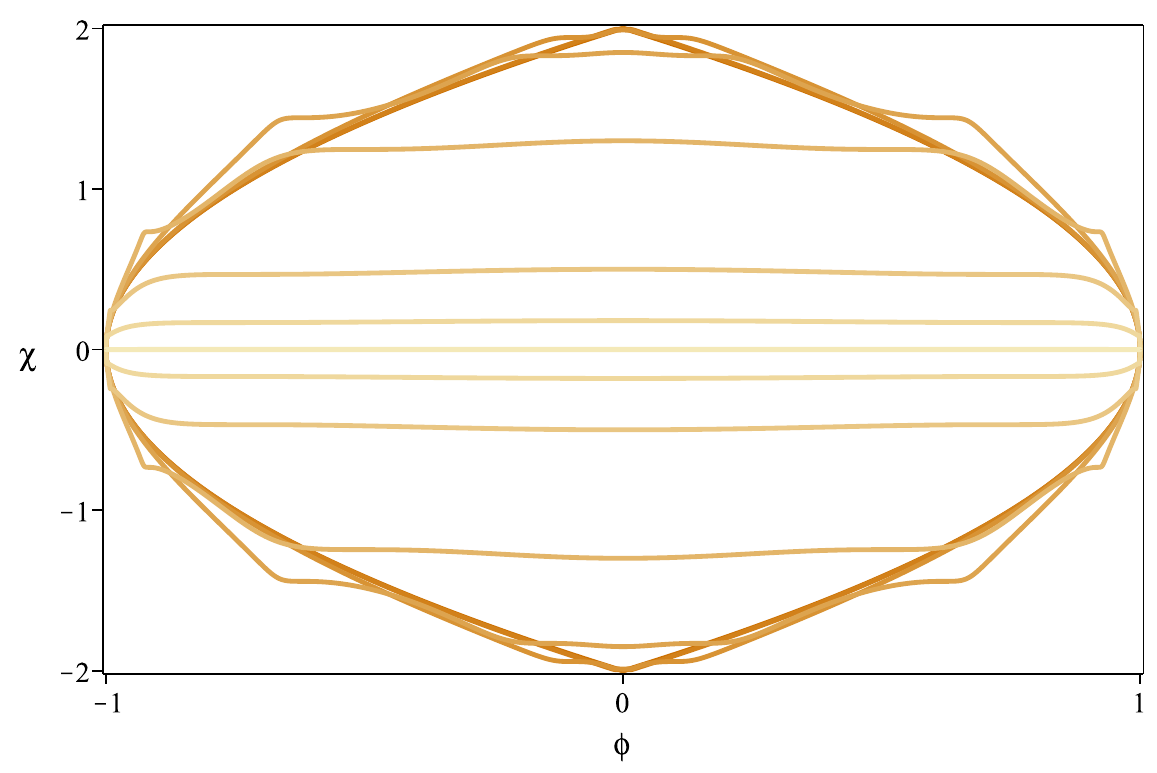}
    \caption{In the top panels, we display the solutions of Eqs.~\eqref{firstfg}, with the boundary conditions $\phi(0)=0$ and $\chi(0)=\chi_0$, for $\phi(x)$ in blue and $\chi(x)$ in red, in which $r=1/4$, $n=2$, $\alpha=1/4$ and $\chi_0=0.001, 0.18, 0.5, 1.3, 1.85, 1.99, 1.999$ and $1.9999$. In the middle panels, we depict the energy density in green without the three smallest values of $\chi_0$. In the bottom panels, we show the orbits in orange. In the left panels, we take $f(\psi)=1/\cos^2(n\pi\psi)$ and $g(\psi) = 1$; in the right ones, we consider $f(\psi)=1$ and $g(\psi)=1/\cos^2(n\pi\psi)$. The colors get lighter as $\chi_0$ decreases.}
    \label{figfg1}
\end{figure}

To illustrate the procedure, we first consider the case in which $f(\psi)=1/\cos^2(n\pi\psi)$ and $g(\psi)=1$. For this, the above first-order equations do not support analytical solutions. By using numerical procedures, we solve the above equations with the boundary conditions $\phi(0)=0$ and $\chi(0)=\chi_0$ and depict them in the top-left panel of Fig.~\ref{figfg1} for $r=1/4$, $n=2$, $\alpha=1/4$ and $\chi_0=0.001, 0.18, 0.5, 1.3, 1.85, 1.99, 1.999$ and $1.9999$. Notice that only $\phi(x)$ gets plateaus at the points in which $f(\psi(x))$ diverges, while $\chi(x)$ only has the intensity of its slope slightly increased at these points, deforming its bell shape (compare it with the standard solutions depicted in the top-left panel of Fig.~\ref{figparsolrho1}). The effect of this modification in the energy density can be seen in the middle-left panel of Fig.~\ref{figfg1}: it attains minima which are not zeroes of $\rho_1(x)$. The orbits were constructed numerically for the above set of solutions; they can be seen in the bottom-left panel of Fig.~\ref{figfg1}. We remark that, contrary to the standard case ($f(\psi)=1$) shown in Fig.~\ref{figorbit}, the orbits may intercept each other due to the explicit dependence on the coordinate in Eq.~\eqref{orbiteq}.

We then interchange the previous functions, that is, we take $f(\psi)=1$ and $g(\psi)=1/\cos^2(n\pi\psi)$. We use numerical procedures to find the solutions with the same set of parameters and boundary conditions as before. They are displayed in the top-right panel of Fig.~\ref{figfg1}. In contrast to previous case, $\chi(x)$ now engender plateaus and $\phi(x)$ has the intensity of its slope slightly decreased. The energy density, however, is quite similar to the previous case, with changes in the depth of the valleys and height of the peaks; it can be seen in the middle-right panel of Fig.~\ref{figfg1}. Using the solutions, we obtain the orbits and show them in the bottom-right panel of Fig.~\ref{figfg1}; notice that intersection points between them are also present here.

\begin{figure}[t!]
    \centering
    \includegraphics[width=4.2cm]{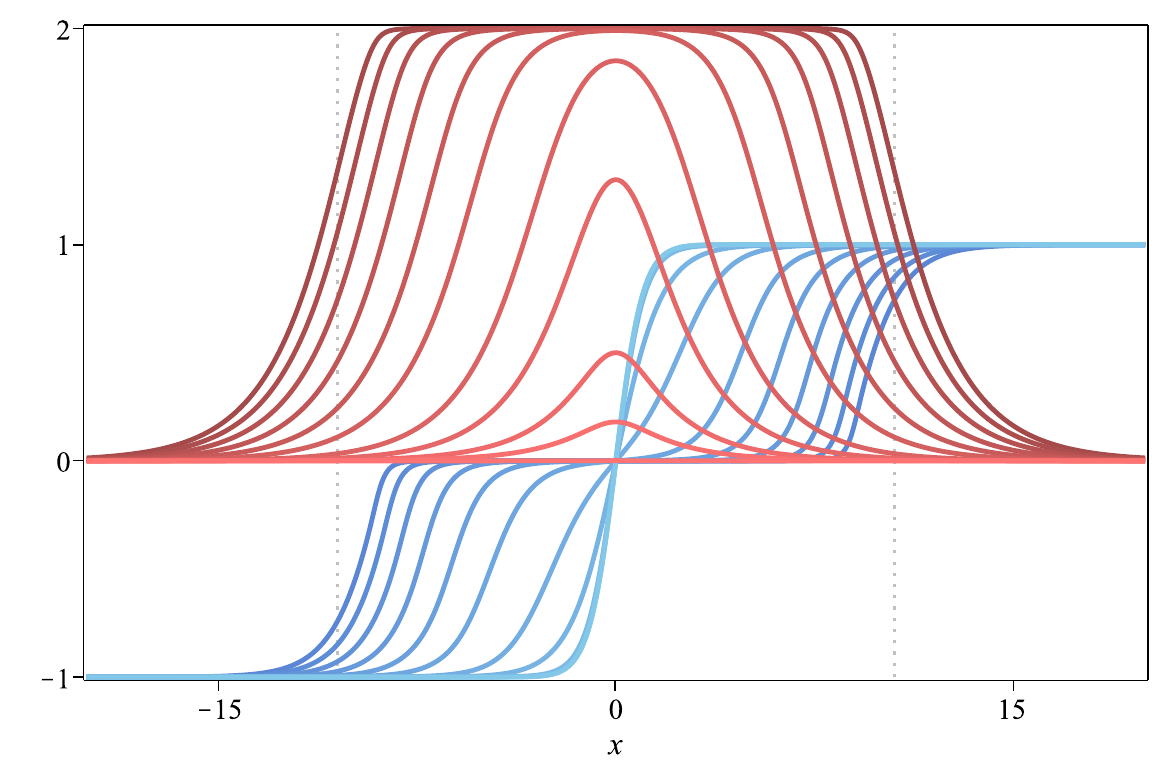}
    \includegraphics[width=4.2cm]{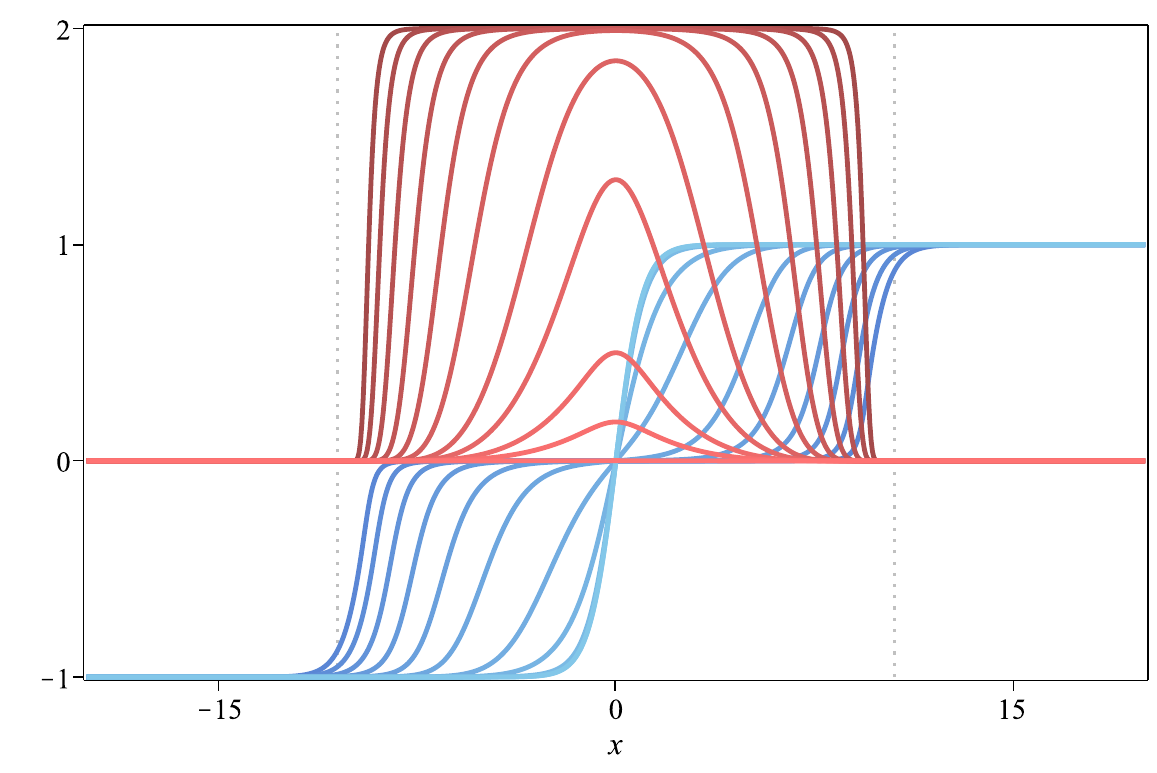}
    \includegraphics[width=4.2cm]{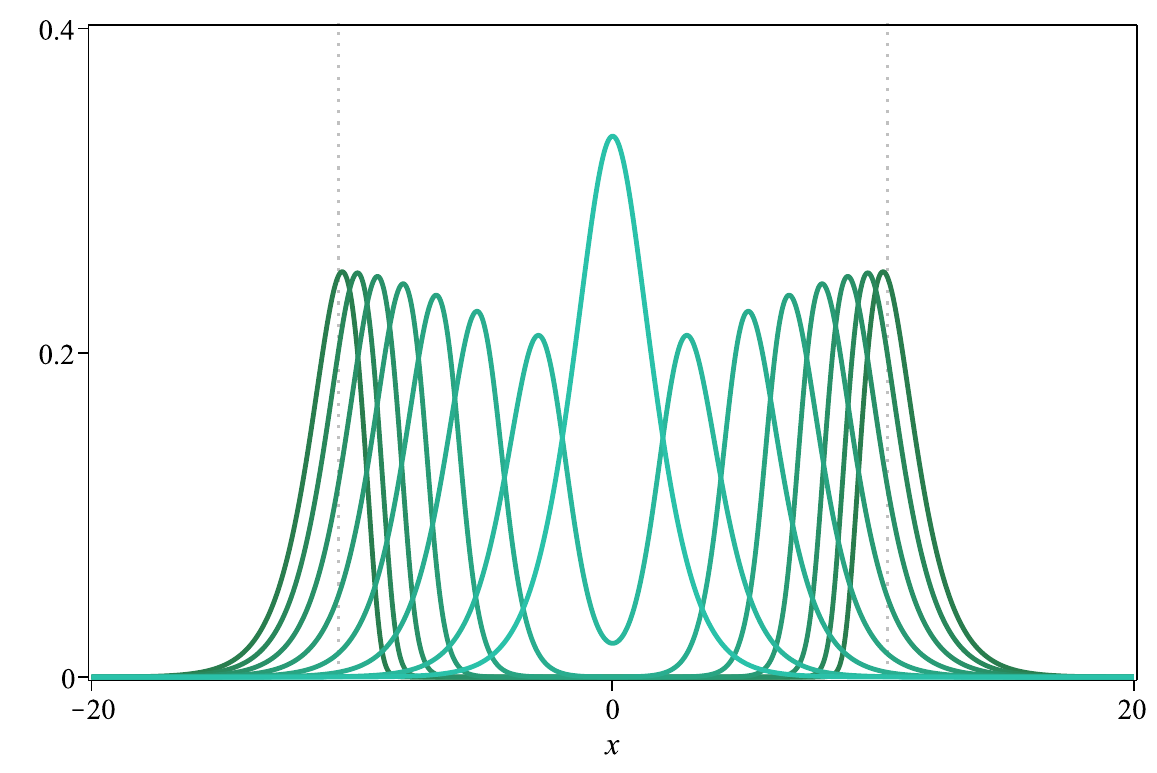}
    \includegraphics[width=4.2cm]{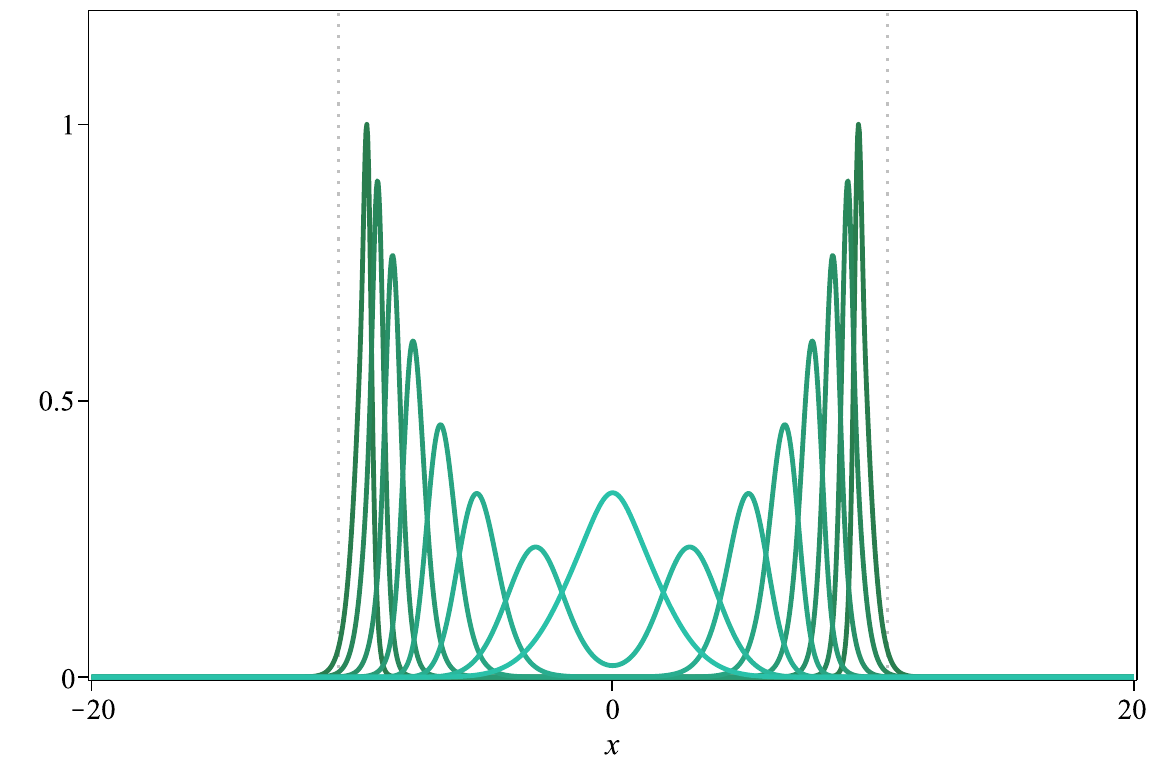}
    \includegraphics[width=4.2cm]{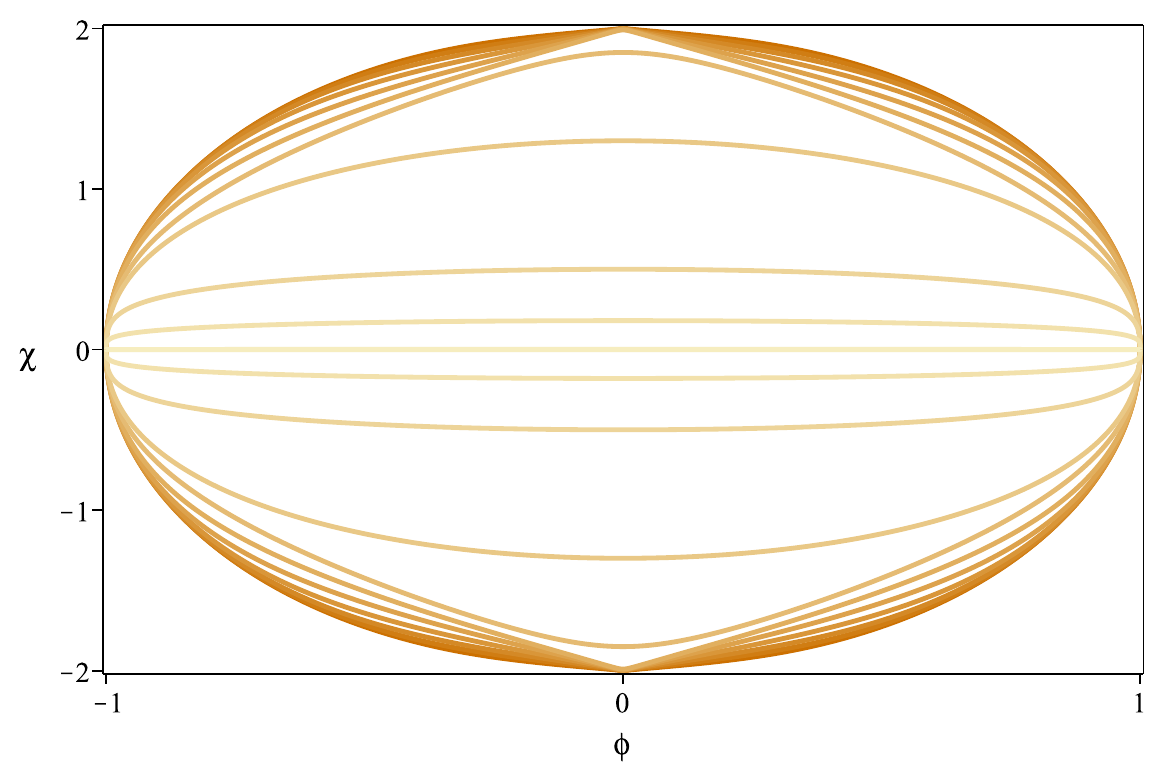}
    \includegraphics[width=4.2cm]{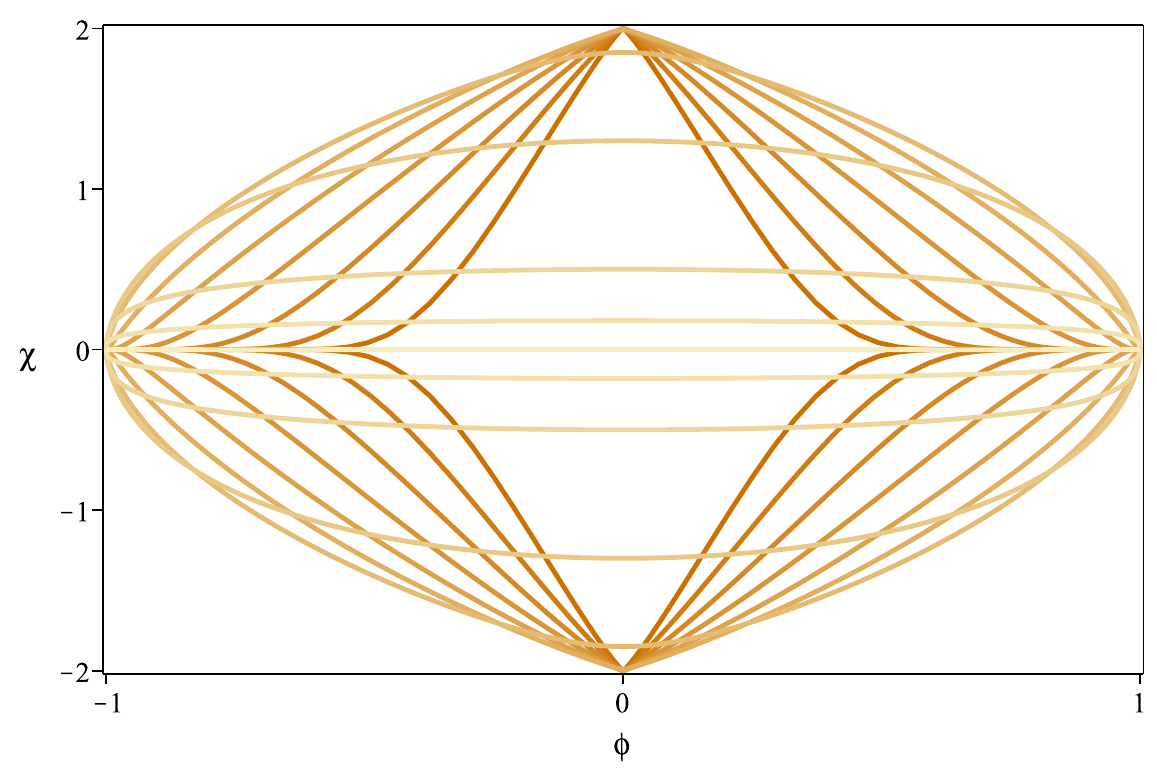}
    \caption{In the top panels, we display the solutions of Eqs.~\eqref{firstfg}, with the boundary conditions $\phi(0)=0$ and $\chi(0)=\chi_0$, for $\phi(x)$ in blue and $\chi(x)$ in red, in which $r=1/4$, $\alpha=1/16$ and $\chi_0\in\Omega$. In the middle panels, we depict the energy density in green without the three smallest values of $\chi_0$. In the bottom panels, we show the orbits in orange. In the left panels, we take $f(\psi)=(1-3\psi^2)^2$ and $g(\psi)=1$; in the right ones, we consider $g(\psi)=(1-3\psi^2)^2$ and $f(\psi)=1$. The dotted vertical lines delimit the compact space $[-x_c,x_c]$, where $x_c=\text{arctanh}(1/\sqrt{3})/\alpha$. The colors get lighter as $\chi_0$ decreases.}
    \label{figfgcomp}
\end{figure}

The solutions in Fig.~\ref{figfg1} present modifications in the core of the structure in comparison with the standard case ($f(\psi)=g(\psi)=1$). We have seen in Sec.~\ref{secf} that one may use the function in Eq.~\eqref{fcomp} with $f(\psi)=g(\psi)$ to compactify the structure. Let us investigate the case in which $f(\psi)\neq g(\psi)$, first considering $f(\psi)=(1-3\psi^2)^2$ and $g(\psi)=1$. We were not able to solve the first-order equations \eqref{firstfg} analytically for the boundary conditions $\phi(0)=0$ and $\chi(0)=\chi_0$. Thus, we make use of numerical methods to plot the solutions in the top-left panel of Fig.~\ref{figfgcomp} for $r=1/4$, $\alpha=1/16$ and $\chi_0\in \Omega$, where $\Omega=\{0.001$, $0.18$, $0.5$, $1.3$, $1.85$, $1.99$, $1.999$, $1.9999$, $1.99999$, $1.999999$, $1.9999999\}$. Interestingly, even though $f(\psi(x))$ presents divergences at $x=\pm x_c$, where $x_c=\text{arctanh}(1/\sqrt{3})/\alpha$, the solutions $\phi(x)$ and $\chi(x)$ are not compact. The associated energy density is displayed in the middle-left panel of Fig.~\ref{figfgcomp}; it has a similar profile of the standard case. The orbits are shown in the bottom-left panel of Fig.~\ref{figfgcomp}.

Finally, we interchange the functions, i.e., take $f(\psi)=1$ and $g(\psi)=(1-3\psi^2)^2$, and solve the first-order equations \eqref{firstfg} for the same set of parameters and boundary conditions of the previous case. In this situation, the solution $\chi(x)$ is compact, connecting the minima of the potential in the space $x\in[-x_c,x_c]$; see the top-right panel of Fig.~\ref{figfgcomp}. The solution $\phi(x)$, on the other hand, has a faster falloff due to the function $g(\psi(x))$, but not enough to compactify the solution. It is worth to highlight that, since one of the solutions is not compact, the energy density is not compact, as it is shown in the middle-right panel of Fig.~\ref{figfgcomp}. By using the solutions, we get the orbits depicted in the bottom-right panel of Fig.~\ref{figfgcomp}.

\section{Model 2}\label{sec2}
In the model described by the Lagrangian density \eqref{lmodel1}, we have considered the $\psi$ to modify the dynamical terms of the fields associated to the BNRT solutions, $\phi$ and $\chi$. We now investigate a distinct possibility, described by
\be\label{lmodel2}
\begin{aligned}
    \mathcal{L} &= \frac12h(\phi,\chi)\partial_\mu\psi\partial^\mu\psi + \frac12\partial_\mu\phi\partial^\mu\phi\\&+\frac12\partial_\mu\chi\partial^\mu\chi - V(\phi,\chi,\psi).
\end{aligned}
\ee
In this case, the BNRT fields, $\phi$ and $\chi$, are the source fields, acting to modify the kinetic term of $\psi$ via the mediator function $h(\phi,\chi)$, as represented below.
\begin{center}
\tikzset{terminal2/.style  = {draw, circle, color=white, text=black, fill=white, minimum size=2em}}
\begin{tikzpicture}[
terminal/.style={ circle,     minimum width=1cm,       minimum height=1cm,      ultra thin, draw=black,
       font=\itshape}
      ]
\matrix[row sep=-0.2cm,column sep=0.2cm] {%
    \node [terminal](p1) {$\phi$};   & & & \\
        &\node [terminal2](p3) {$h(\phi,\chi)$};& & \node [terminal](p4) {$\psi$}; \\
    \node [terminal](p2) {$\chi$}; & & &\\
}; 

\draw[gray]   (p1) edge [->,>=stealth,shorten <=2pt, shorten >=2pt,thick] (p4)
        (p2)  edge [->,>=stealth,shorten <=2pt, shorten >=2pt, thick] (p4);
      
\end{tikzpicture}
\end{center}
We suppose that the mediator function is non negative in order to avoid the presence of negative contributions in the energy density.

The equations of motion associated to the source fields are
\bes
\bal
&\partial_\mu\partial^\mu\phi -\frac{1}{2} h_\phi \partial_\mu\psi\partial^\mu\psi + V_\phi =0, \\
&\partial_\mu\partial^\mu\chi -\frac{1}{2} h_\chi \partial_\mu\psi\partial^\mu\psi + V_\chi =0.
\eal
\ees
For $\psi$, we have the following equation of motion
\be
\partial_\mu\left(h(\phi,\chi)\,\partial^\mu \phi\right)+V_\phi=0.
\ee
Since we are interested in studying localized structures, we take static configurations, such that the above equations of motion take the form
\be
\phi'' -\frac{1}{2} h_\phi {\psi'}^2 = V_\phi,\qquad\chi'' -\frac{1}{2} h_\chi {\psi'}^2 = V_\chi
\ee
and
\be
\left(h(\phi,\chi)\,\psi'\right)'=V_\psi.
\ee
In this model, the energy density is written as
\be
\rho = \frac{1}{2}h(\phi,\chi){\psi'}^2 +   \frac12{\phi'}^2+\frac12{\chi'}^2 + V(\phi,\chi,\psi).
\ee
Notice that the equations of motion are of second order, with couplings between the fields. To get a first-order framework, we develop the Bogomol'nyi procedure rewriting the energy density as
\be
\begin{aligned}
    \rho &= \frac12h(\phi,\chi)\left(\psi'\mp \frac{W_\psi}{h(\phi,\chi)}\right)^2 + \frac12\!\left(\phi'\mp W_\phi \right)^2\\
    &+ \frac12\!\left(\chi'\mp W_\chi \right)^2 +V -\frac12\!\left(\frac{W^2_\psi}{h(\phi,\chi)} +W^2_\phi + W^2_\chi \right) \pm W',
\end{aligned}
\ee
in which we have introduced the auxiliary function $W(\phi,\chi,\psi)$. By considering that the potential is non negative, with the form
\be
V(\phi,\chi,\psi) = \frac{W^2_\psi}{2h(\phi,\chi)} +\frac12 W^2_\phi + \frac12 W^2_\chi,
\ee
the energy can be written as in Eq.~\eqref{eb1}, being minimized to $E=E_B$ if the first-order equations $\phi'= \pm W_\phi$, $\chi'= \pm W_\chi$ and $\psi'= \pm W_\psi/h(\phi(x),\chi(x))$ are satisfied. We only use the equations with the upper sign, as they are related to the ones with lower sign by the change $x\to-x$. In principle, $W$ can couple $\phi$, $\chi$ and $\psi$. An interesting situation appears if the auxiliary function is decomposed as in Eq.~\eqref{wpq}, because we can define the geometrical coordinate $\zeta$ via the expression
\be\label{zeta}
\frac{d\zeta}{dx} = \frac{1}{h(\phi(x),\chi(x))} \implies\zeta = \int\frac{dx}{h(\phi(x),\chi(x))}.
\ee
Notice that $\zeta$ depends on the solutions of the source fields. The first-order equation associated to $\psi$ then becomes
\be\label{fopsizeta}
\psi_\zeta = Q_\psi,
\ee
where $\psi_\zeta=d\psi/d\zeta$. To solve this equation, one needs to calculate the source fields with the equations
\be\label{fophichi2}
\phi'= P_\phi \quad\text{and}\quad \chi'= P_\chi.
\ee
In general, the model supports an infinite number of pairs $(\phi(x),\chi(x))$ that satisfy the above first-order equations. Since the coordinate $\zeta$ in Eq.~\eqref{zeta} depends on these source fields, for a given $h(\phi,\chi)$ there may be several distinct solutions $\psi(x)$.

The above first-order equations allow us to write the energy density as $\rho = \rho_1 + \rho_2$, where
\bes
\bal
\rho_1(x) &= \frac{Q^2_\psi}{h(\phi(x),\chi(x))},\\
\rho_2(x) &= P^2_\phi+P^2_\chi
\eal
\ees

To illustrate our procedure, we consider the source fields $\phi$ and $\chi$ described by the BNRT model in Eqs.~\eqref{pbnrt}. We focus on the solutions \eqref{solc}, which connect the horizontal minima of the diamond and depend on the orbit parameter $c$,  having the associated energy density in Eq.~\eqref{rhoc}, with change $\rho_1\to\rho_2$. As we shall see, this approach is interesting, as the source fields always connect the same minima, but the orbit parameter induces modifications in the third field. We consider the field $\psi$ to be governed by
\be\label{qpsi42}
Q(\psi) = \psi-\frac13\psi^3.
\ee
From Eq.~\eqref{fopsizeta}, we get the solution and energy density
\be\label{psirhozeta}
\psi(x) = \tanh(\zeta)\quad\text{and}\quad \rho_1(x) = \frac{\sech^4(\zeta)}{h(\phi(x),\chi(x))}.
\ee
The coordinate $\zeta$ must be obtained from Eq.~\eqref{zeta}. Notice that, since $h(\phi,\chi)$ depends on both functions, obtaining analytical expressions becomes a harder task when compared to the situation investigated in Sec.~\ref{secf}. However, we have found some cases with analytical results. Let us consider
\be
h(\phi,\chi) = \frac{1}{A\phi^2+B\chi^2},
\ee
\begin{figure}[t!]
    \centering
    \includegraphics[width=4.2cm]{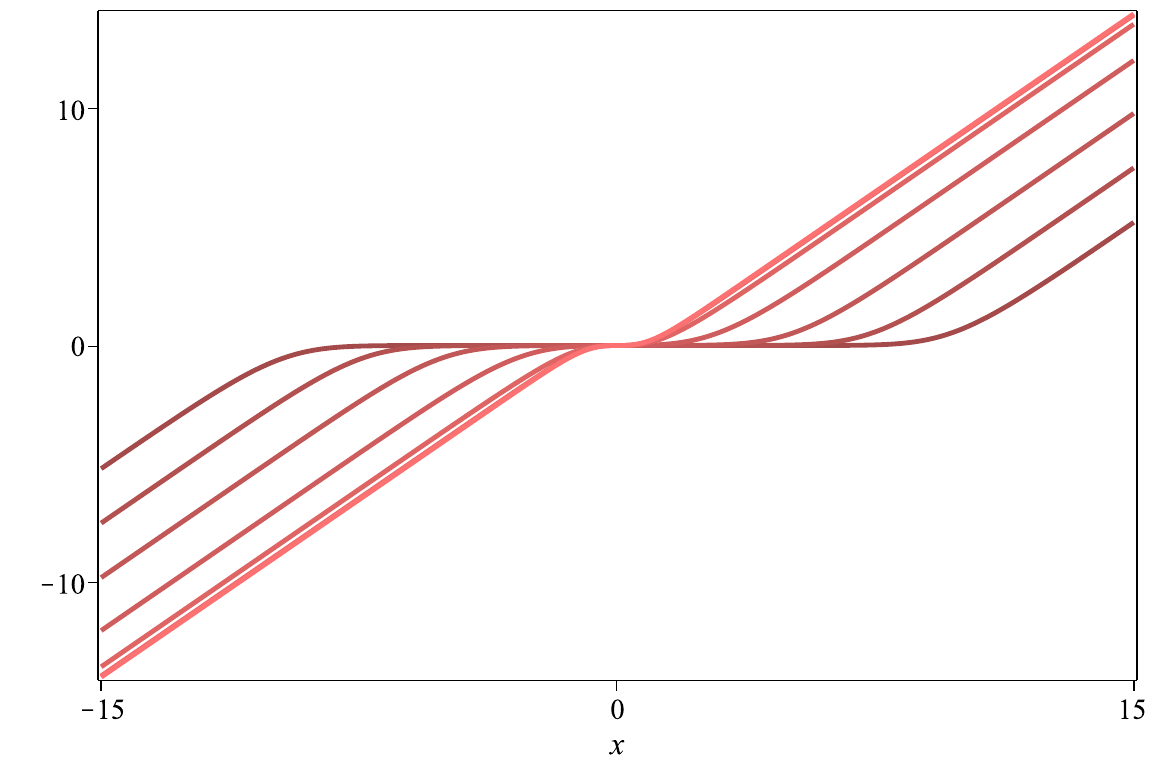}
    \includegraphics[width=4.2cm]{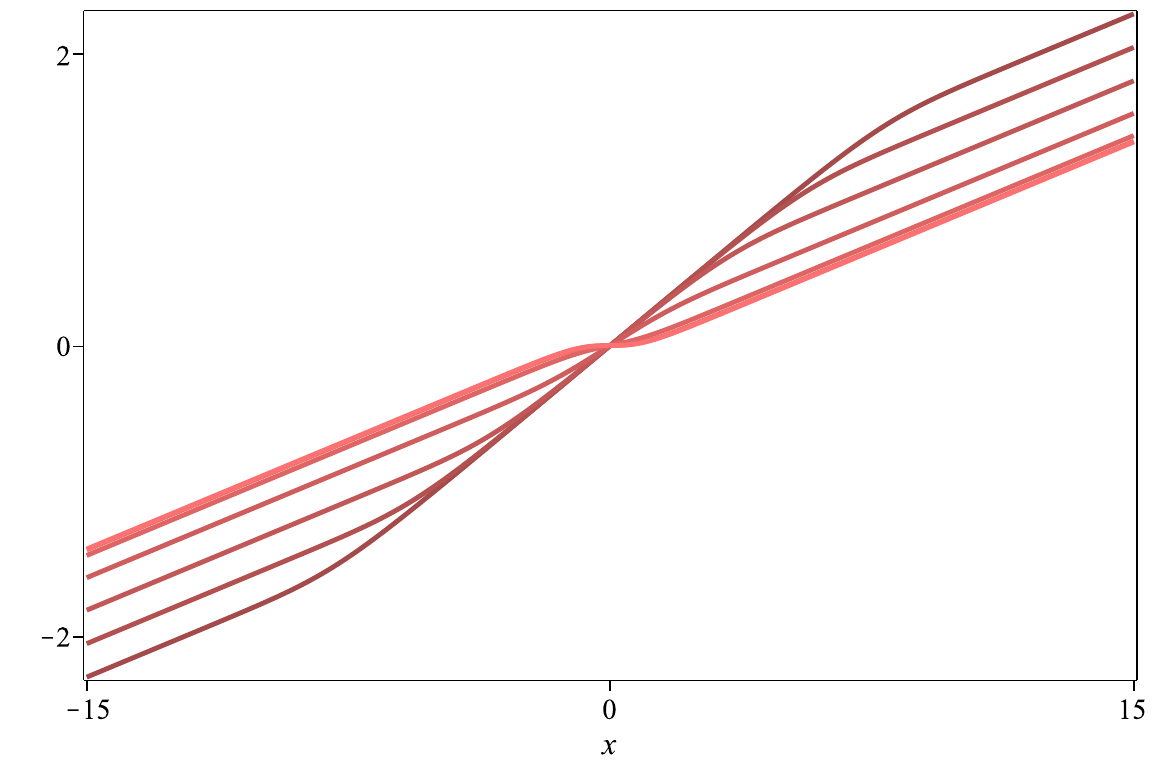}
    \includegraphics[width=4.2cm]{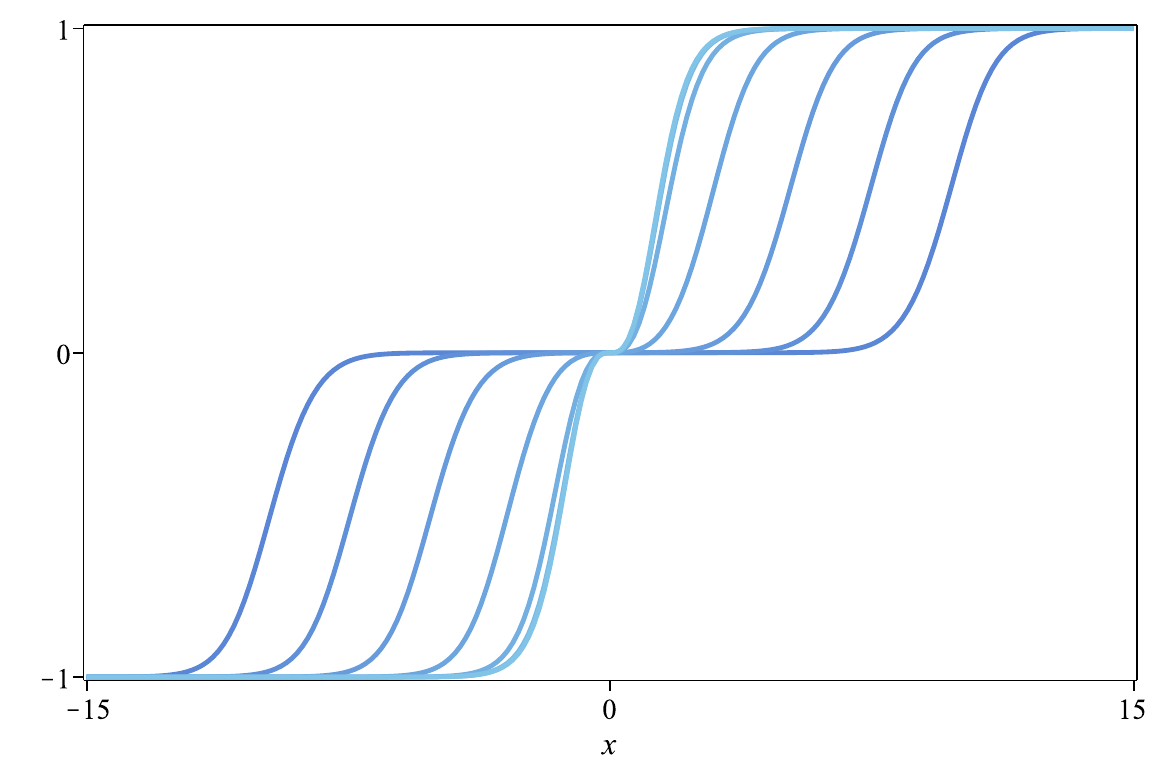}
    \includegraphics[width=4.2cm]{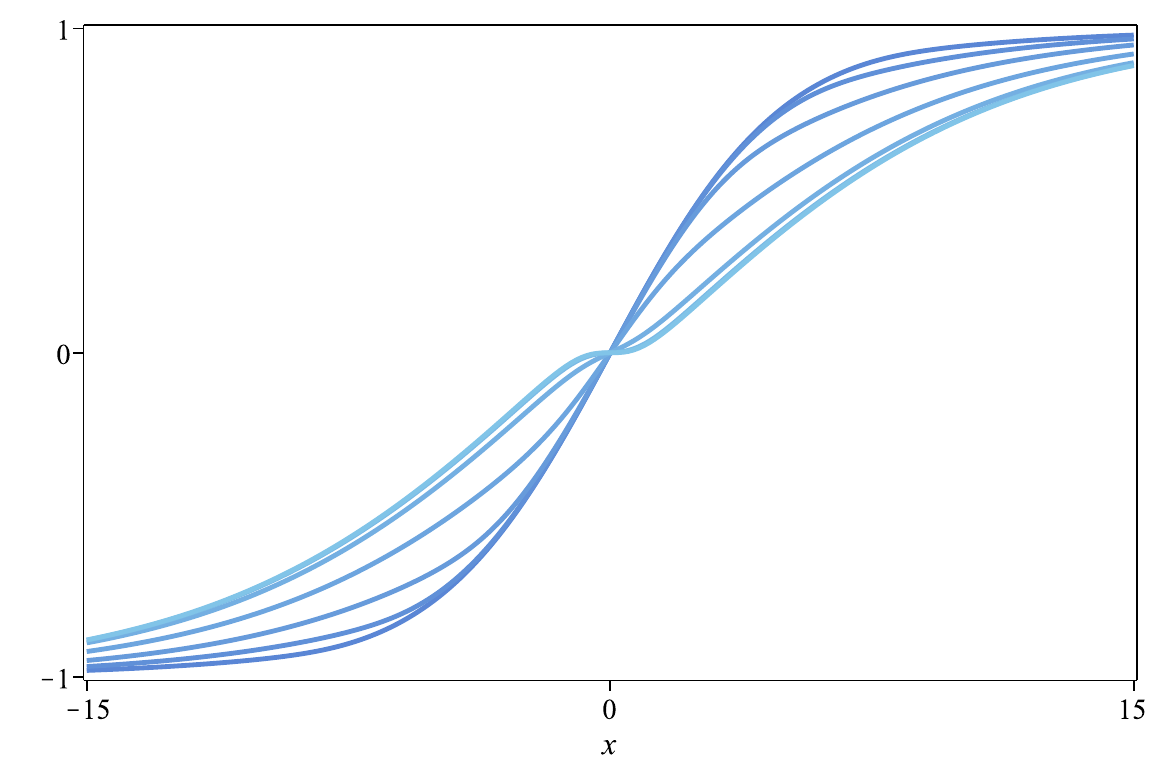}
    \includegraphics[width=4.2cm]{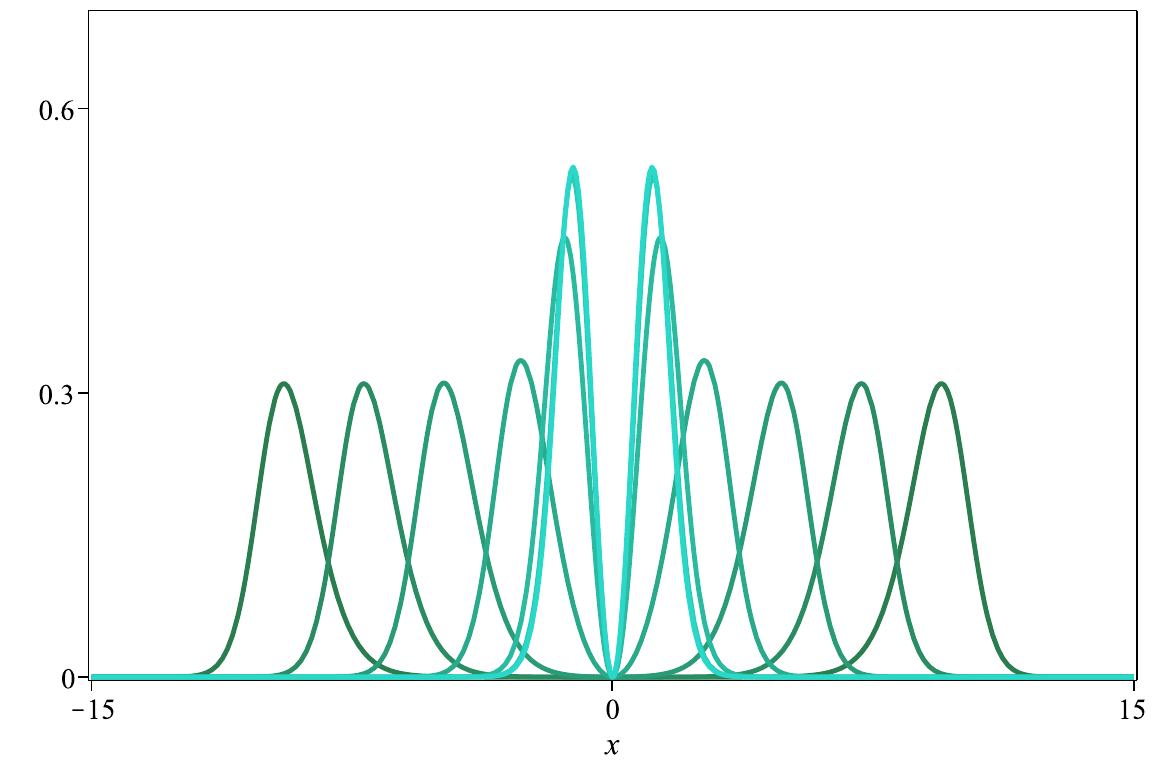}
    \includegraphics[width=4.2cm]{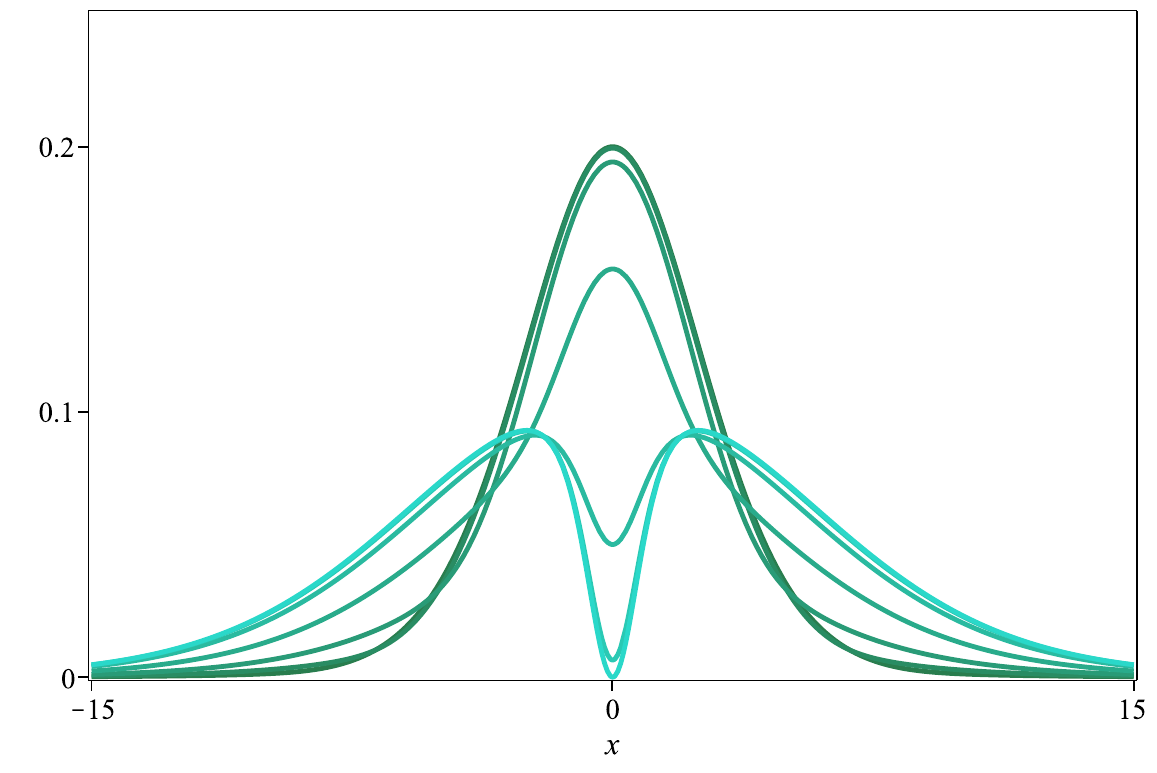}
    \caption{The coordinate $\zeta(x)$ in Eq.~\eqref{zeta1} (top, red color), its associated solution $\psi(x)$ (middle, blue color) and energy density $\rho_1(x)$ (bottom, green) in Eq.~\eqref{psirhozeta} for $c= 10^k$, with $k=-4,-3,\ldots 1$, and $c\to\infty$. In the left panels we take $A=1$ and $B=0$, and in the right ones we consider $A=1/10$ and $B=1/20$. The colors gets lighter as $c$ increases.}
    \label{figh1}
\end{figure}
where $A$ and $B$ are, \emph{a priori}, non-negative parameters to ensure the positiveness of the above function. In this case, Eq.~\eqref{zeta} leads us to
\begin{widetext}
\bes\label{zeta1}
\begin{align}
    \zeta(x) & = Ax-\dfrac{2Ac\tanh\left(\dfrac{x}{2}\right)}{(c-1)\tanh^2\left(\dfrac{x}{2}\right)+c+1}+\dfrac{2(4B-A)}{\sqrt{1-c^2}}\text{arctanh}\left(\sqrt{\dfrac{1-c}{1+c}}\tanh\left(\dfrac{x}{2}\right)\right),\quad\text{for}\quad c<1,\\
    \zeta(x) & = Ax+2(2B-A)\tanh\left(\dfrac{x}{2}\right), \quad\text{for}\quad c=1,\\
    \zeta(x) & = Ax-\dfrac{2Ac\tanh\left(\dfrac{x}{2}\right)}{(c-1)\tanh^2\left(\dfrac{x}{2}\right)+c+1}+\dfrac{2(4B-A)}{\sqrt{c^2-1}}\arctan\left(\sqrt{\dfrac{c-1}{c+1}}\tanh\left(\dfrac{x}{2}\right)\right),\quad\text{for}\quad c>1.
\end{align}
\ees
\end{widetext}
To get $\zeta(x)$ with infinite range, we must impose $A\neq0$. Near the origin, for $x\approx0$, we have $\zeta(x)\approx 4Bx/(c+1)$. Asymptotically,  one can show that $\zeta(x)$ goes to a straight line. In Fig.~\ref{figh1}, we depict the above function, and its respective solution and energy density \eqref{psirhozeta}. In the case where $A=1$ and $B=0$, displayed in the left panels, we see that $h(\phi,\chi)$ only acts to deform the geometry of the solution around the origin, keeping the asymptotic behavior of $\zeta(x)$ as a straight line. This induces a plateau in the solution $\psi(x)$ and a valley in the energy density at the center of the structure for all $c$. If one consider a non-vanishing $B$, as shown in the right panels, the parameter $c$ that comes from the source fields acts to modify the slope of the solution at its center, inducing a plateau around the origin as it gets larger. In this situation, $c$ also modifies the peak of the energy density, which goes down and becomes a valley for $c>c_v$, where $c_v\approx1.072$ was obtained numerically.

As commented in Sec.~\ref{secf}, in order to compactify the structure, one must consider functions that support zeroes. Following this insight, we take the $\chi$-dependent mediator function
\be\label{hcomp}
h(\chi) = \left(1-\chi^2\right)^2.
\ee
One may include a parameter to control the strength of $\chi^2$, with a term in the form $a \chi^2$. The only requirement to get the modifications in which we are interested is that $a<2$, so we have taken $a=1$ without loss of generality. This function only vanishes at points where $\chi^2=1$. However, since $\chi(x)$ is given as in Eq.~\eqref{solc}, the above equation becomes
\be
h(\chi(x)) = \left(\frac{3-c\cosh(x)}{1+c\cosh(x)}\right)^2.
\ee
Thus, this function does not engender zeroes for all values $c$, which is the parameter associated to the orbits; see Eqs.~\eqref{bnrtorbitr} and \eqref{solc}. We have found three possibilities regarding the zeroes of $h(\chi)$. It presents i) two symmetric zeroes, given by $x=\pm x_c$, where $x_c\equiv\text{arccosh}\,(3/c)$, for $c<3$; ii) a single zero ($x=0$) for $c=3$; iii) no zero for $c>3$.

\begin{figure}[t!]
    \centering
    \includegraphics[width=4.2cm]{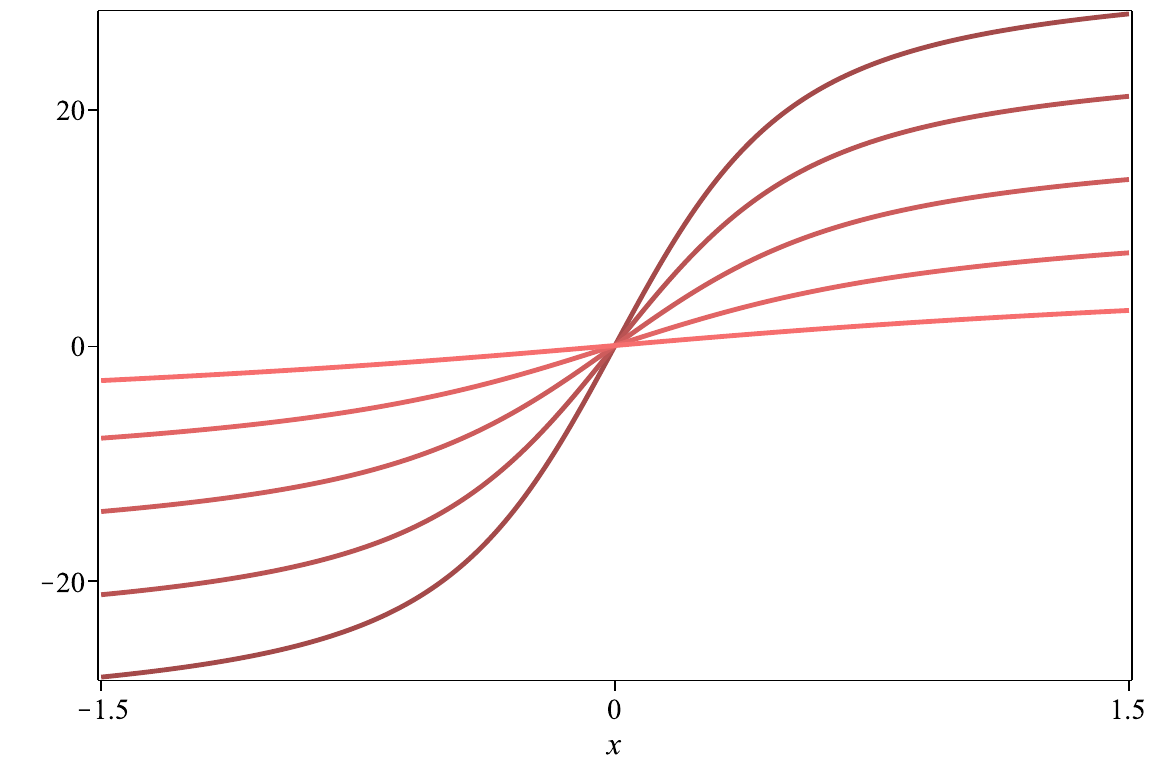}
    \includegraphics[width=4.2cm]{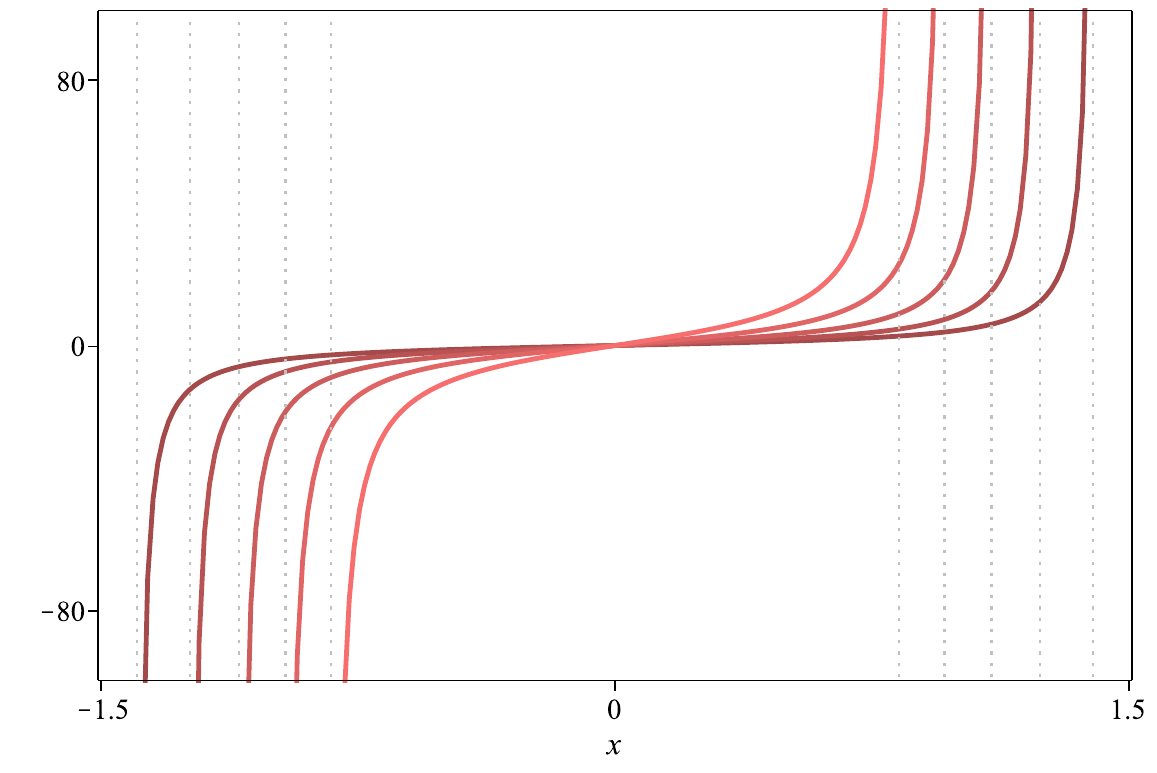}
    \includegraphics[width=4.2cm]{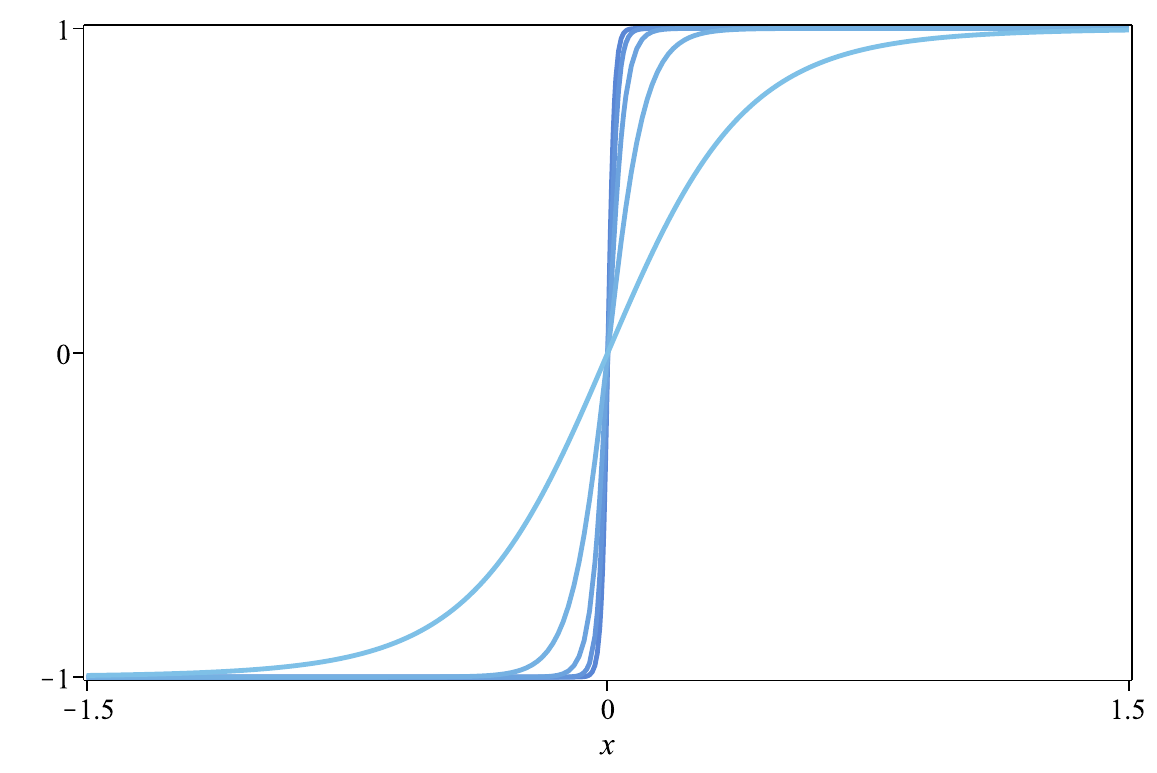}
    \includegraphics[width=4.2cm]{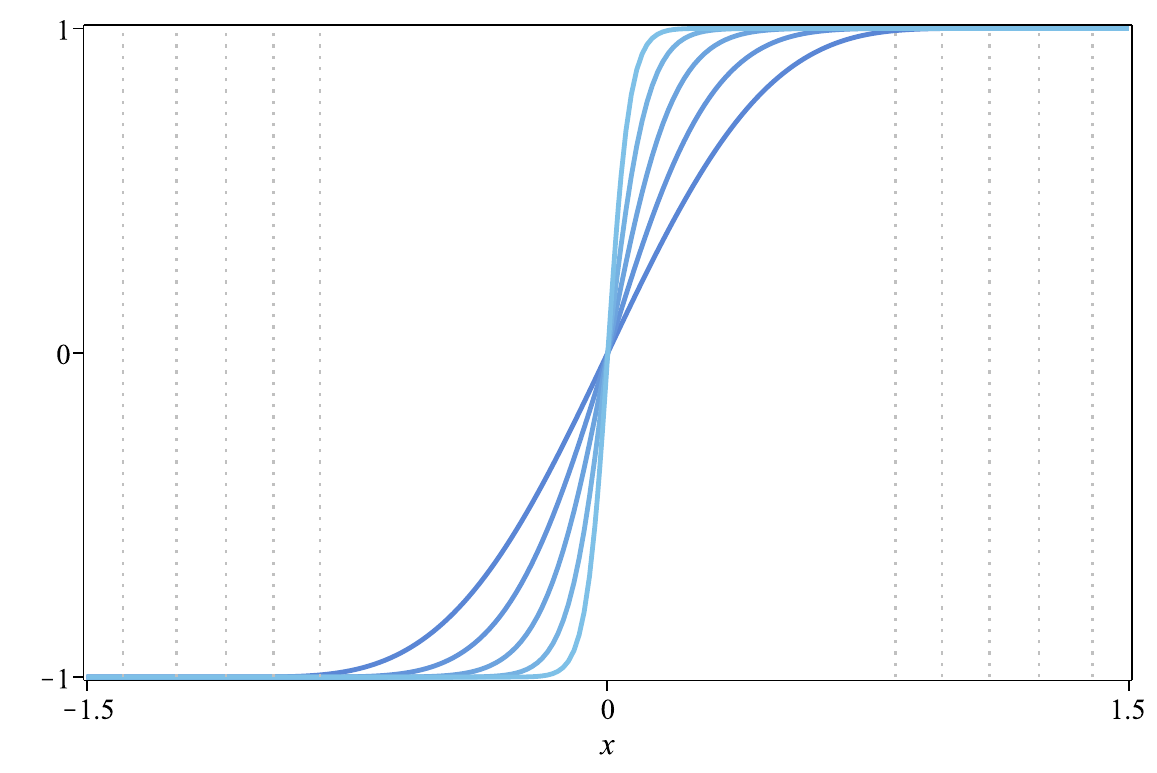}
    \includegraphics[width=4.2cm]{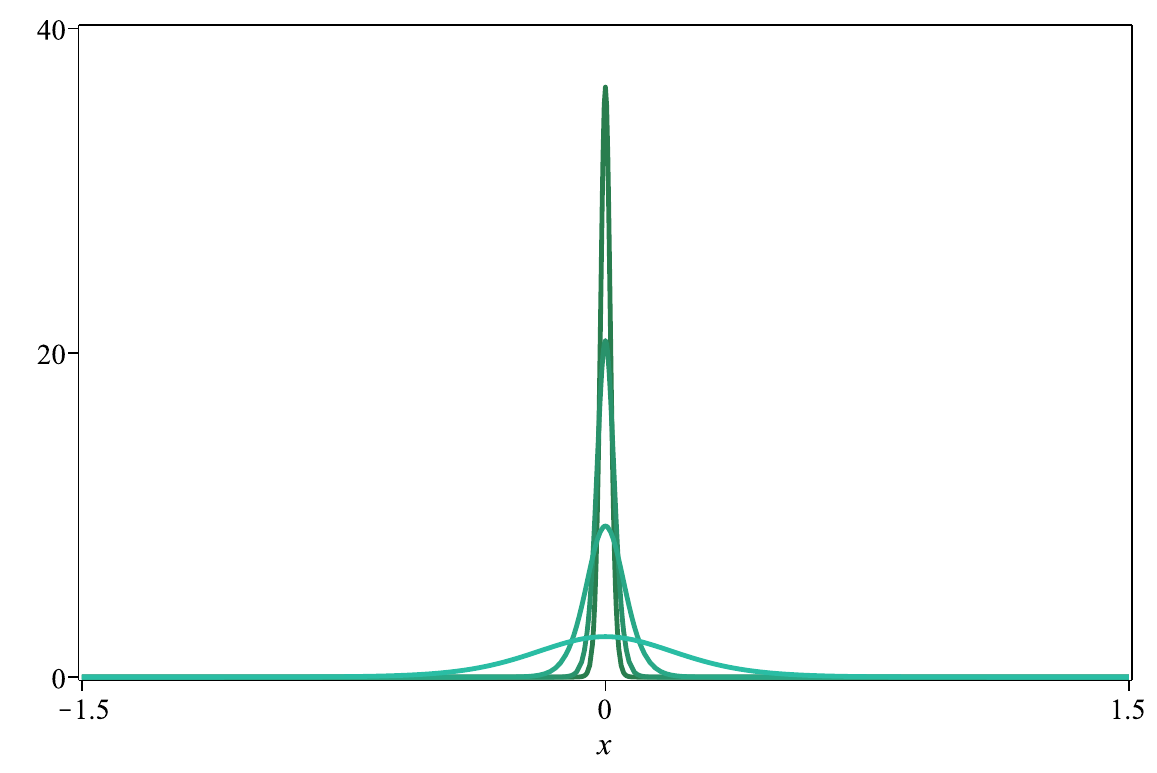}
    \includegraphics[width=4.2cm]{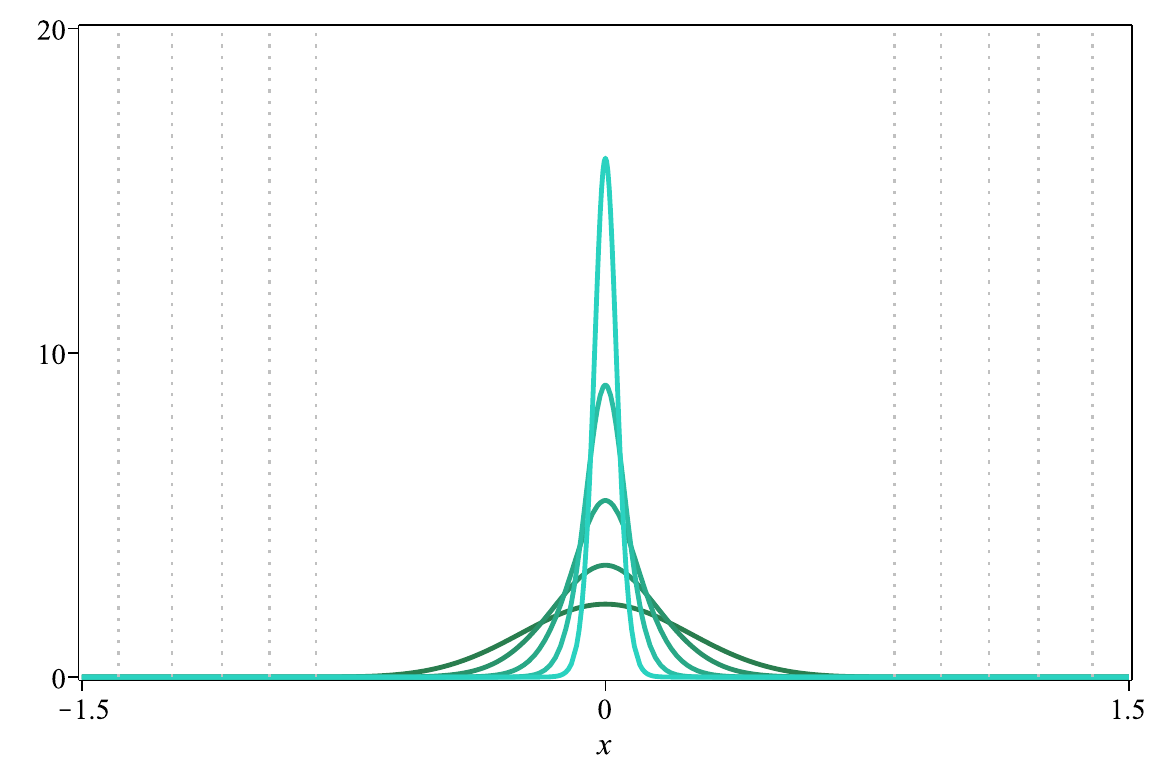}
    \caption{The coordinate $\zeta(x)$ in Eq.~\eqref{zeta2a} (top left, red color) and  its associated solution $\psi(x) $(middle left, blue color) and energy density $\rho_1(x)$ (bottom left, green color) in Eq.~\eqref{psirhozeta} for $c=3.63, 3.795, 4.125, 4.95$ and $9.9$ (we removed $c=3.63$ in the energy density for convenience). In the right panels, we display these quantities for the coordinate $\zeta(x)$ in Eq.~\eqref{zetacomp} for $c=1.4, 1.6, 1.8, 2$ and $2.2$. The dotted-vertical lines delimits the compact space, $|x|\leq x_c$. The colors gets lighter as $c$ increases.}
    \label{figh2}
\end{figure}

The case without zeroes ($c>3$) does not lead to significant modifications in the function $h(\chi)$; its associated geometric coordinate \eqref{zeta} is
\begin{widetext}
\be\label{zeta2a}
    \zeta(x) = x+\dfrac{32c\tanh\left(\dfrac{x}{2}\right)}{(c^2-9)\left((c+3)\tanh^2\left(\dfrac{x}{2}\right)+c-3\right)}+\dfrac{16(c^2-3)}{(c^2-9)^{3/2}}\,\text{arctan}\left(\sqrt{\dfrac{c+3}{c-3}}\tanh\left(\dfrac{x}{2}\right)\right).
\ee
For the case with two zeroes ($c<3$), the geometric coordinate connects $-\infty$ to $+\infty$ in a limited space, $|x|<x_c$. It has the form
\be\label{zetacomp}
\zeta(x) = \begin{cases}x-\dfrac{32c\tanh\left(\dfrac{x}{2}\right)}{(9-c^2)\left((c+3)\tanh^2\left(\dfrac{x}{2}\right)+c-3\right)} -\dfrac{16(3-c^2)}{(9-c^2)^{3/2}}\,\text{arctanh}\left(\sqrt{\dfrac{3+c}{3-c}}\tanh\left(\dfrac{x}{2}\right)\right),& |x|<x_c,\\
\text{sgn}(x)\,\infty,&|x|>x_c,
\end{cases}
\ee
\end{widetext}
with $\text{sgn}\,(x)$ denoting the sign function. Near the points $x\approx \pm x_c$ inside the compact space, the geometric coordinate behaves as $\zeta(x)\propto 1/(x-x_c)$. In Fig.~\ref{figh2}, we depict the coordinates in Eqs.~\eqref{zeta2a} (left panels) and \eqref{zetacomp} (right panels), and their associated solutions $\psi(x)$ and energy density $\rho_1(x)$ in Eq.~\eqref{psirhozeta} for several values of $c$. By looking at the left panels, one can see that, as $c$ gets near $3$ from the right, the thickness of the structure gets smaller, leading to a thin solution with a tall peak in the energy density. As $c$ gets larger and larger, approaching the straight-line orbit, the geometric coordinate tends to behave as $\zeta(x)\approx x$, such that the usual hyperbolic tangent profile is recovered in the solution. In the right panels, we can see that the structure is compactified in the space $|x|\leq x_c$, with $c$ controlling its width. In this situation, as $c$ gets larger, approaching $3$, the thickness of the structure decreases, similarly to the case in the left panels. However, contrary to the situation depicted in the left panels, since $c<3$, the solutions are compact.

When the mediator function \eqref{hcomp} has a single zero ($c=3$), $\zeta(x)$ becomes simpler:
\be\label{zetahc}\zeta(x) = \begin{cases}
 x-\dfrac{8}{27}\,\dfrac{1-6\tanh^2\left(\dfrac{x}{2}\right)}{\tanh^3\left(\dfrac{x}{2}\right)}, & x>0,\\
-\infty, & x<0.
\end{cases}
\ee
At the right neighborhood of $x=0$, the above expression becomes $\zeta(x)\propto-1/x^3$, showing how the divergence emerges in the function. This divergence at the origin makes the solution in Eq.~\eqref{psirhozeta} become a half compact, with its left tail compactified, since $\psi(x)=-1$ for $x<0$. This also occurs in the energy density, which engender the usual lump-like shape.

It is worth commenting that the analysis in this section can also be done considering the source fields given by other solutions of the BNRT model. For instance, one can consider \eqref{sol116}, which is associated to the orbit that connects neighbor minima in the diamond. This makes the mediator functions to attain different profiles, which may lead to other interesting results.

\section{Final Remarks}\label{secend}
In this work, we have investigated two classes of three-field models. First, we have considered the case in which a single source field, $\psi$,  acts to modify the fields $\phi$ and $\chi$ via the mediator functions $f(\psi)$ and $g(\psi)$ as in Eq.~\eqref{lmodel1}. We have developed a first-order formalism based on energy minimization to show that, if the auxiliary function has the form \eqref{wpq}, the first-order equation that governs the source field is decoupled. We then have studied the case in which $\psi$ acts on the solutions of BNRT model, which is described by \eqref{pbnrt}. Interestingly, if the mediator functions are equal, $f(\psi)=g(\psi)$, the orbits formed by $\phi$ and $\chi$ are the same of the standard case ($f(\psi)=g(\psi)=1)$. This fact allowed us to define a geometric coordinate, $\xi$, associated to the source field, which replaces $x$ in the standard solution. In this direction, we have investigated the case in which the mediator functions contain a parameter that dictates the number of their divergent points and gives rise to symmetric plateaus outside the origin in the solution. We have also presented the possibility of including a plateau at the origin. To modify the tail of the BNRT solutions, we have considered a mediator functions with two zeroes that induce the compactification of the structure. In this situation, the geometric coordinate ranges from $-\infty$ to $\infty$ in a limited space.

Models in which the mediator functions are different, $f(\psi)\neq g(\psi)$, have also been investigated. In this case, the problem becomes harder, as one cannot redefine the geometric coordinate to obtain the solutions in terms of the ones associated to the standard case. Also, the equation \eqref{orbiteq} becomes explicitly dependent on the coordinate, so it is impossible to find a general orbit to decouple the first-order equations. We have used numerical methods to calculate the solutions for specific boundary conditions and construct the orbit. To illustrate this situation, we have presented four examples. The results show that the solutions may present stark differences when the functions $f(\psi)$ and $g(\psi)$ are interchanged. For instance, in the case $f(\psi)=(1-3\psi^2)^2$ and $g(\psi)=1$ there is no compact solution. On the other hand, for $f(\psi)=1$ and $g(\psi)=(1-3\psi^2)^2$, the solution $\chi(x)$ is compact, even though $\phi(x)$ is not.

The possibility of having the solutions of the BNRT model \eqref{pbnrt} acting as source fields to modify a single scalar field via the mediator function has also been investigated with the Lagrangian density \eqref{lmodel2}. By considering that the scalar field is driven by \eqref{qpsi42}, we have shown that one can obtain the same solutions of the standard model with argument defined by a novel geometric coordinate, $\zeta$. In this case, since there is an infinite number of solutions connecting the horizontal minima of the BNRT model, the same mediator function may lead to different modifications. For instance, by using \eqref{hcomp}, we have shown that, depending on the orbit parameter, one may get non-compact, half-compact or compact structures.

We believe that the procedure presented in the current work may have applications in some issues of interest in Physics. For instance, one may consider to use the source fields and mediator functions to induce modifications in other structures, such as vortices \cite{n1,n2,n3} and monopoles \cite{m1,m2,m3}. Since the solutions present distinct features in their core and/or tail, they may also be useful in the study of solutions in other known models \cite{izquierdosol}, collisions of kinks \cite{scat1,scat2,scat3,scat4,scat5,scat6} and domain wall networks \cite{net1,net2,net3}. In the curved spacetime, it may be used in the study of scalar field solutions around static black holes \cite{moreira}, in wormhole systems \cite{wormhole} or in other geometries \cite{morris}.

\vspace{1cm}
{\noindent\textbf{Data availability statement}: This manuscript has no associated data.}

\acknowledgements{This work is supported by the Brazilian agency Conselho Nacional de Desenvolvimento Cient\'ifico e Tecnol\'ogico (CNPq), grants Nos. 306151/2022-7 (MAM) and 310994/2021-7 (RM). It is also supported by Paraiba State Research Foundation (FAPESQ-PB), grants Nos. 0003/2019 (RM) and 0015/2019 (MAM).}


\end{document}